%% file: sim_fgas.tex
\documentclass[useAMS,twocolumn,usenatbib]{mn2e}
\usepackage{times,psfig,epsfig} 
\usepackage{rotating, graphicx}
\usepackage{color}

\newcommand{\xmm}{{\it XMM-Newton}}

\def \xmm {\hbox{\it XMM-Newton}}

\newcommand{\nr}{NR}
\newcommand{\w}{CSF}
\newcommand{\agn}{AGN}

\pssilent 

\newcommand{\be}{\begin{equation}}
\newcommand{\ee}{\end{equation}}
\newcommand{\ba}{\begin{eqnarray}}
\newcommand{\ea}{\end{eqnarray}}
\newcommand{\brr}{\begin{array}}
\newcommand{\err}{\end{array}}
\newcommand{\bc}{\begin{center}}
\newcommand{\ec}{\end{center}}

\newcommand{\msun}{\,h^{-1}M_\odot}

\newcommand{\vel}{\,{\rm km\,s^{-1}}}

\newcommand{\mincir}{\raise
  -2.truept\hbox{\rlap{\hbox{$\sim$}}\raise5.truept \hbox{$<$}\ }}
\newcommand{\magcir}{\raise  
 -2.truept\hbox{\rlap{\hbox{$\sim$}}\raise5.truept \hbox{$>$}\ }}
\newcommand{\siml}{\raise
  -2.truept\hbox{\rlap{\hbox{$\sim$}}\raise5.truept \hbox{$<$}\ }}
\newcommand{\simg}{\raise
  -2.truept\hbox{\rlap{\hbox{$\sim$}}\raise5.truept \hbox{$>$}\ }}

\title[Baryon Census in Hydrodynamical Simulations of Galaxy Clusters]
{Baryon Census in Hydrodynamical Simulations of Galaxy Clusters}

\author[S. Planelles et al.]
{S. Planelles$^{1,2}$\thanks{e-mail: susana.planelles@oats.inaf.it} , S. Borgani$^{1,2,3}$, 
K. Dolag$^{4,5}$, S. Ettori$^{6}$, D. Fabjan$^{3,7,8}$, 
G. Murante$^{2}$, 
\newauthor
L. Tornatore$^{1}$  \\~\\
\footnotesize 
$^1$ Astronomy Unit, Department of Physics, University of Trieste, via Tiepolo 11, I-34131 Trieste, Italy\\
$^2$ INAF, Osservatorio Astronomico di Trieste, via Tiepolo 11, I-34131 Trieste, Italy\\ 
$^3$ INFN -- National Institute for Nuclear Physics, Trieste,Italy\\ 
$^4$ Universit\"atssternwarte M\"unchen, M\"unchen, Germany\\
$^5$ Max-Planck-Institut f\"ur Astrophysik, Garching, Germany\\
$^6$ INAF, Osservatorio Astronomico di Bologna, via Ranzani 1, I-40127 Bologna, Italy\\
$^7$ SPACE-SI, Slovenian Centre of Excellence for Space 
          Sciences and Technologies, A$\check{s}$ker$\check{c}$eva 12, 1000 Ljubljana, Slovenia \\
$^8$ Faculty of Mathematics and Physics, University of Ljubljana, Jadranska 19, 1000 Ljubljana, Slovenia \\
 }

\begin{document}
\maketitle 

\begin{abstract}
  We carry out an analysis of a set of cosmological SPH hydrodynamical
  simulations of galaxy clusters and groups aimed at studying the
  total baryon budget in clusters, and how this budget is shared
  between the hot diffuse component and the stellar component. Using
  the TreePM+SPH {\tt GADGET-3} code, we carried out one set of
  non--radiative simulations, and two sets of simulations including
  radiative cooling, star formation and feedback from supernovae (SN),
  one of which also accounting for the effect of feedback from active
  galactic nuclei (AGN). The analysis is carried out with the twofold
  aim of studying the implication of stellar and hot gas content on
  the relative role played by SN and AGN feedback, and to calibrate
  the cluster baryon fraction and its evolution as a cosmological
  tool. 
  With respect to previous similar analysis, the simulations
  used in this study provide us with a sufficient statistics of massive
  objects and including an efficient AGN feedback.
  We find that both radiative simulation sets predict a trend of
  stellar mass fraction with cluster mass that tends to be weaker than the
  observed one. However this tension depends on the particular set of
  observational data considered. Including the effect of AGN feedback
  alleviates this tension on the stellar mass and predicts values of
  the hot gas mass fraction and total baryon fraction to be in closer
  agreement with observational results. We further compute the ratio
  between the cluster baryon content and the cosmic baryon fraction,
  $Y_b$, as a function of cluster-centric radius and redshift.  At
  $R_{500}$ we find for massive clusters with $M_{500}>2\times
  10^{14}\msun$ that $Y_b$ is nearly independent of the physical
  processes included and characterized by a negligible redshift
  evolution:  $Y_{b,500}=0.85\pm 0.03$ with the error accounting for
  the intrinsic r.m.s. scatter within the set of simulated
  clusters. At smaller radii, $R_{2500}$, the typical value of $Y_b$
  slightly decreases, by an amount that depends on the physics
  included in the simulations, while its scatter increases by about a
  factor of two. These results have interesting implications for the
  cosmological applications of the baryon fraction in clusters.
\end{abstract} 
 
\begin{keywords}  
cosmology: miscellaneous -- methods: numerical -- galaxies: cluster: general 
-- X-ray: galaxies. 
\end{keywords}

\section{Introduction}

Galaxy clusters, which stem from the collapse of density fluctuations
involving comoving scales of tens of Mpc, are the largest
gravitationally bound structures in the Universe.  Since they are
relatively well isolated systems, knowledge of their baryon content is
a key ingredient to understand the physics of these objects and their
use as cosmological probes \citep[see][for recent reviews]{Allen2011,
  kravtsov_borgani12}. The most massive galaxy clusters are of
particular cosmological interest since their baryon content is
expected to trace accurately the baryon content of the Universe.  In
the absence of dissipation, the ratio of baryonic-to-total mass in
clusters should closely match the ratio of the cosmological parameters
measured from the cosmic microwave background (CMB) and, therefore,
$M_b/M_{tot} \sim \Omega_b/\Omega_m$ \citep{White1993, Evrard1997,
  Ettori2003, Allen2008}.
  
However, contrary to expectations, measurements of the baryon mass
fraction in nearby clusters from optical and X-ray observations have
posed a challenge to this fundamental assumption since they have
reported smaller values than expected \citep[e.g.,][]{Ettori2003,
  Biviano2006, McCarthy2007} together with a possible intriguing trend
with cluster mass \citep[e.g.,][]{Lin2003, Lin2012} for a broad mass
range of systems.  With the precise measurement of the universal
baryon fraction, 
$f_b\equiv \Omega_b/\Omega_m=0.167\pm0.004$, from the {\it
  Wilkinson Microwave Anisotropy Probe}
\citep[WMAP-7;][]{Komatsu2011}, these discrepancies have gained in
physical importance and claim for different explanations: physical
processes which lower $f_\mathrm{b}$ in clusters relative to the
universal fraction \citep[see e.g.,][]{Bialek2001, He2006},
significant undetected baryon components by standard X-ray and/or
optical techniques \citep[see][]{Ettori2003, Lin2004}, or a systematic
underestimate of $\Omega_\mathrm{m}$ by WMAP \citep{McCarthy2007}.

In this sense, the correct determination of the gas mass fraction may
be crucial. In fact, studies of the individual baryon components have
shown that the stellar and gas mass fractions within
R$_{500}$\footnote{R$_\mathrm{\Delta}$ ($\Delta$=2500, 500, 200) is
the radius within which the mass density of a group/cluster is equal
to $\Delta$ times the critical density ($\rho_{c}$) of the
Universe. Correspondingly,
M$_\mathrm{\Delta}=\Delta\,\rho_{c}(z)\,(4\,\pi/3)R_{\Delta}^3$ is
the mass inside R$_\mathrm{\Delta}$.} exhibit opposite behaviours as
a function of the total system mass. In particular, clusters have a
higher gas mass fraction than groups \citep[e.g.,][]{Vikhlinin2006,
Arnaud2007}, but a lower stellar mass fraction \citep{Lin2003}. This
has been interpreted as a difference in the star formation efficiency
between groups and clusters \citep[e.g.,][]{David1990, Lin2003,
Lagana2008}.

On the other hand, the mass dependence of the gas fraction and the
discrepancy between the baryon mass fraction in groups/clusters and
the WMAP value can be understood in terms of non--gravitational
processes.  In this regard, AGN heating, which can drive the gas
outside the potential wells of host halos, can explain the lack of gas
within $R_{500}$ in groups.  Therefore, groups also appear as critical
systems to assess the universality of the baryon fraction and to
understand complex physical processes affecting both the gas and the
stellar components.

The baryonic mass content of clusters consists of stars in cluster
galaxies (satellite galaxies plus the brightest cluster galaxy or
BCG), intra-cluster light (ICL, stars that are not bound to cluster
galaxies), and the hot intra-cluster medium (ICM) whose mass exceeds
the mass of the former two stellar components by a factor of $\sim 6$.
Therefore, to reliably estimate the cosmological parameters from the
baryonic-to-total mass ratio one should address all the baryonic
components and not only the gas mass.  In fact, to explain the
discrepancy between the observed baryon fraction and the cosmic value,
the ICL is suggested to be one of the most important forms of missing
baryons, accounting for $6-22$ per cent of the total cluster light in
the r--band \citep[e.g.,][]{Gonzalez2007}, or even more in cluster
mergers \citep[e.g.,][]{Pierini2008}.  Cosmological simulations report
that the ICL accounts for up to $\simeq 60$ per cent of the total of
stars \citep[e.g.,][]{Murante2004, Murante2007, Puchwein2010}.
However, although this large fraction of ICL is sufficient to explain
the discrepancy with the cosmic value, it seems to be in contradiction
with observations.  On one hand, this suggests that other mechanisms,
like gas expulsion by AGN heating, may also be important to account
for the missing baryons. On the other hand, one should also bear in
mind that the methods to identify ICL in simulations and in
observations are quite different, the former being often based on
criteria of gravitational boundness of star particles, while the
latter being based on criteria of surface brightness limits
\citep[e.g.][]{rudick_etal06}.

The combination of robust measurements of the baryonic mass fraction
in clusters from X-ray observations together with a determination of
$\Omega_{\rm b}$ from CMB data or big-bang nucleosynthesis
calculations and a constraint on the Hubble constant, can therefore be
used to measure $\Omega_{\rm m}$ \citep[e.g.,][]{White1991, White1993,
  Evrard1997, Allen2002, Ettori2003, Lin2003, Allen2003, Allen2004,
  Allen2008}. This method, remarkably simple and robust in terms of
its underlying assumptions, currently provides strong 
constraints on $\Omega_{\rm m}$.  On the other hand, measurements of
the apparent redshift evolution of the cluster X-ray gas mass
fraction, hereafter $f_{g}$, can also be used to constrain the
geometry of the Universe \citep{Sasaki1996}, its acceleration
\citep{Pen1997} and, therefore, the dark energy equation of state
\citep[e.g.,][]{Ettori2003, ettori_etal09, Allen2002, Allen2003,
  Allen2004, LaRoque2006, Allen2008, Allen2011}.  This diagnostic
exploits the dependence of the $f_{g}$ measurements (derived from the
observed X-ray gas temperature and density profiles) on the assumed
distances to the clusters, $f_{\rm g}\propto d^{1.5}$, and relies on
cluster baryon fractions being roughly universal and non-evolving over
the redshift range where they can be observed (typically $z<1$).
Therefore, like Type Ia supernovae, massive clusters serve as standard
calibration sources that test the expansion history of the universe.
   
In this regard, early simulations by \cite{Eke1998} have been used to
calibrate the depletion factor, i.e., the ratio by which the baryon
fraction measured at $R_{2500}$ is depleted with respect to the
universal mean \citep[see, for instance,][]{Allen2008}.  These
non-radiative simulations indicate that, within the virial radius, the
baryon fraction in clusters provides a measure of the universal mean
that is only slightly biased low by $\sim 10$ per cent.  The magnitude of
this bias might change if additional physics is included in the
simulations.  Therefore, it is necessary and extremely useful to
deepen in the analysis of how different implementations of baryonic
physics can affect the value of this bias, its evolution with
redshift, its radial dependence within clusters, or some statistics
with massive galaxy clusters.
   
Using non-radiative hydrodynamical simulations the expectation is
that, for the largest ($kT >5$\,keV), dynamically relaxed clusters and
for measurement radii beyond the innermost core ($r > R_{2500}$),
$f_{g}$ should be approximately constant with redshift
\citep[e.g.,][]{Eke1998, Crain2007}. However, possible systematic
variations of $f_{g}$ with redshift can be accounted for in a
straightforward manner if the allowed range of such variations is
constrained by numerical simulations or other complementary data
\citep{Eke1998, Bialek2001, Muanwong2002, Borgani2004, Kay2004,
  Ettori2004, Ettori2006, Kravtsov2005, Nagai2007}.

Therefore, it is clear that understanding the baryon mass fraction and
its mass and redshift dependence is a crucial issue to understand
better astrophysics in galaxy clusters, e.g., the origin of the ICL
\citep[e.g.,][]{Pierini2008}, star-formation history
\citep[e.g.,][]{Fritz2010}, metal-enrichment history
\citep[e.g.,][]{Kapferer2009}, the dynamical history of galaxy
clusters, and the use of these systems to constrain the cosmological
parameters \citep[e.g.,][]{Allen2011}.

The purpose of the present work is to use a set of hydrodynamical
simulations of galaxy clusters, characterized by different physical
processes, to study how the fraction and spatial distribution of
baryons, as contributed both by the stellar component and by the hot
X-ray emitting gas, are affected by the physical conditions within
clusters and how these results compare with observations.  We also
analyse how the different baryonic depletions depend on redshift,
baryonic physics, and cluster radius, determining, therefore, some
implications for the constraints on cosmological parameters derived
from gas mass fractions within clusters.  In this regard, our analysis
extends previous analyses of the baryon fraction in cluster
simulations which included only non--radiative physics
\citep[e.g.,][]{Evrard1990, Metzler1994, Navarro1995, Lubin1996,
 Eke1998, Frenk1999, Mohr1999, Bialek2001}, 
 the processes of cooling and star formation 
  \citep[e.g.,][]{Muanwong2002, Kay2004, Ettori2004,
  Kravtsov2005,Ettori2006, Deboni2011, sembolini_etal12}, 
  and the effect of AGN feedback 
  \citep[e.g.,][]{Puchwein2008, Puchwein2010, fabjan_etal10, 
  Young2011, Battaglia2012}.
  In addition, with respect to previous similar analysis, the simulations
  presented in this work provide us with a sufficient statistics of massive
  objects and including an efficient AGN feedback.

The paper is organized as follows: in Section~2, we describe our
dataset of simulated galaxy clusters and the different physical
processes considered in re-simulating them; in Section~3, we present
the results obtained from this set of simulations on the baryon, gas
and stellar mass fractions as a function of cluster mass and we
compare these results with different observational data sets; in
Section~4 we calibrate the different baryonic depletions and analyse
their dependences on redshift, baryonic physics and cluster
radius; and finally, in Section~5, we summarize and discuss our
findings.

\section{The simulated clusters}
\label{sec:simulations}

\subsection{Initial conditions}
Our sample of simulated clusters and groups are obtained from 29
Lagrangian regions, centred around as many massive halos identified
within a large-volume, low-resolution N-body cosmological simulation
\citep[see][for details]{Bonafede2011}.  The parent Dark Matter (DM)
simulation followed $1024^3$ DM particles within a box having a
comoving side of 1 $h^{-1}$ Gpc, with $h$ the Hubble constant in units
of 100~km~s$^{-1}$~Mpc$^{-1}$.  The cosmological model assumed is a
flat $\Lambda$CDM one, with $\Omega_{\rm{m}} = 0.24$ for the matter
density parameter, $\Omega_{\rm{b}} = 0.04$ for the contribution of          
baryons, $H_0=$~72~km~s$^{-1}$~Mpc$^{-1}$ for the present-day Hubble
constant, $n_{\rm{s}}=$~0.96 for the primordial spectral index and
$\sigma_8 = 0.8$ for the normalisation of the power spectrum.  Within
each Lagrangian region we increased the mass resolution and added the
relevant high-frequency modes of the power spectrum, following the
zoomed initial condition (ZIC) technique \citep{Tormen1997}. Outside
these regions, particles of mass increasing with distance from the
target halo are used, so as to keep a correct description of the large
scale tidal field.  Each high-resolution Lagrangian region is shaped
in such a way that no low-resolution particle contaminates the central
halo at $z = 0$ at least out to 5 virial radii\footnote{The virial
  radius, $R_{vir}$, is defined as the radius encompassing the overdensity of
  virialization, as predicted by the spherical collapse model
  \citep[e.g.,][]{Eke1996}.}. As a result, each
region is sufficiently large to contain more than one interesting halo
with no contaminants within its virial radius.

Initial conditions have been generated by adding a gas component only
in the high--resolution region, by splitting each particle into two,
one representing DM and another representing the gas component, with a
mass ratio such to reproduce the cosmic baryon fraction. The mass of
each DM particle is $m_{\rm{DM}} = 8.47 \cdot 10^8 \, \msun$ and the
initial mass of each gas particle is $m_{\rm{gas}} = 1.53 \cdot 10^8
\, \msun$.

\subsection{The simulation models}
All the simulations have been carried out with the TreePM--SPH
{\footnotesize {\sc GADGET-3}} code, a more efficient version of the
previous {\footnotesize {\sc GADGET-2}} code \citep{springel05}. In
the high--resolution region gravitational force is computed by
adopting a Plummer-equivalent softening length of $\epsilon =
5\,\rm{h}^{-1}$ kpc in physical units below $z = 2$, while being kept
fixed in comoving units at higher redshift \citep[see][for an
  analysis of the effect of softening on radiative simulations of
  galaxy clusters]{Borgani2006}. As for the hydrodynamic
forces, we assume the minimum value attainable by the SPH smoothing
length of the B--spline kernel to be half of the corresponding value
of the gravitational softening length.

Besides a set of non--radiative hydrodynamic simulations
({\tt \nr} hereafter), we carried out two sets of radiative simulations.

Radiative cooling rates are computed by following the
same procedure presented by \cite{wiersma_etal09}. We account for 
the presence of the cosmic microwave background (CMB) and of UV/X--ray
background radiation from quasars and galaxies, as computed by
\cite{haardt_madau01}. The contributions to cooling from each one of
eleven elements (H, He, C, N, O, Ne, Mg, Si, S, Ca, Fe) have been
pre--computed using the publicly available {\footnotesize {\sc
CLOUDY}} photo--ionisation code \citep{ferland_etal98} for an
optically thin gas in (photo--)ionisation equilibrium.  This
procedure to compute cooling rates avoids the assumptions of
collisional ionisation equilibrium and solar relative abundances,
which were underlying in the prescription based on cooling
rates from \cite{Sutherland1993}, that we adopted in previous
simulations \citep[e.g.][]{tornatore_etal07, fabjan_etal10}.

A first set of radiative simulations includes star formation and the
effect of feedback triggered by supernova (SN) explosions ({\tt \w}
hereafter). As for the star formation model, gas particles
above a given threshold density are treated as multiphase, so as to
provide a sub--resolution description of the interstellar medium,
according to the model originally described by
\cite{springel_hernquist03}. Within each multiphase gas particle, a
cold and a hot-phase coexist in pressure equilibrium, with the cold
phase providing the reservoir of star formation. The production of
heavy elements is described by accounting for the contributions from
SN-II, SN-Ia and low and intermediate mass stars, as described by
\cite{tornatore_etal07}. Stars of different mass, distributed
according to a Chabrier IMF \citep{chabrier03}, release metals over
the time-scale determined by the mass-dependent life-times of
\cite{padovani_matteucci93}.  Kinetic feedback contributed by SN-II is
implemented according to the model by \cite{springel_hernquist03}: a
multi-phase star particle is assigned a probability to be uploaded in
galactic outflows, which is proportional to its star formation rate.
In the {\tt \w} simulation set we assume $\rm{v}_{\rm{w}} = 500\vel$ for
the wind velocity, while assuming a mass--upload rate that is two
times the value of the star formation rate of a given particle.

Another set of radiative simulations is carried out by including the
same physical processes as in the {\tt \w} case, with a lower wind
velocity of $\rm{v}_{\rm{w}} = 350\vel$, but also including the effect
of AGN feedback ({\tt \agn} set, hereafter). The model for AGN
feedback is based on the original implementation presented by
\cite{springel_etal05} (SMH), with feedback energy released as a
result of gas accretion onto
supermassive black holes (BH). In this AGN
model, BHs are described as sink particles, which grow their mass by gas
accretion and merging with other BHs. Gas accretion proceeds at a
Bondi rate, and is limited by the Eddington rate.
Once the accretion
rate is computed for each BH particle, a stochastic criterion is used
to select the surrounding gas particles to be accreted. Unlike in SMH,
in which a selected gas particle contributes to accretion with all its
mass, we included the possibility for a gas particle to accrete only
with a slice of its mass, which corresponds to 1/4 of its original
mass. In this way, each gas particle can contribute with up to four
``generations'' of BH accretion events, thus providing a more
continuous description of the accretion process.

Eddington-limited Bondi accretion produces a radiated energy which
corresponds to a fraction $\epsilon_r= 0.1$ of the rest-mass energy of
the accreted gas, which is determined by the radiation efficiency
parameter $\epsilon_r$. The BH mass is correspondingly decreased by
this amount. A fraction of this radiated energy is thermally coupled
to the surrounding gas. We use $\epsilon_f = 0.1$ for this feedback
efficiency, which increases by a factor of 4 when accretion
enters in the quiescent ``radio'' mode and drops below 
one-hundredth of the limiting Eddington rate
\citep[e.g.][]{sijacki_etal07, fabjan_etal10}.

 Additionally,
  we introduced some technical modifications of the original
  implementation, which will be briefly described here. One difference
  with respect to the original SMH implementation concerns the seeding
  of BH particles.  In the SMH model, BH particles are seeded in a
  halo whenever it first reaches a minimum (total) friends-of-friends
  (FoF) halo mass, such as to guarantee that a halo is resolved with
  at least 50 DM particles.  In order to guarantee that BHs are seeded
  only in halos where star formation took place, in our implementation
  we look for FoF groups in the distribution of star particles, by
  grouping them with a linking length of about 0.05 times the mean
  separation of the DM particles. This linking length is thus smaller
  than that, 0.15--0.20, originally used, to identify virialised
  halos. In the simulations presented here, a minimum mass of $4\times
  10^{10}\msun$ is assumed for a FoF group of star particles to be
  seeded with a BH particle.  Furthermore, we locate seeded BHs at the
  potential minimum of the FoF group, instead of at the density
  maximum, as originally implemented by SMH. We also enforce a more
  strict momentum conservation during gas accretion and BH
  mergings. In this way a BH particle remains within the host galaxy,
  when it becomes a satellite of a larger halo. In the original SMH
  implementation, BHs were forced to remain within the host galaxy by
  pinning them to the position of the particle found having the
  minimum value of the potential among all the particles lying within
  the SPH smoothing length compute at the BH position. We verified
  that an aside effect of this criterion is that, due to the
  relatively large values of SPH smoothing lengths, a BH can be removed
  from the host galaxy whenever it becomes a satellite, and is
  spuriously merged into the BH hosted in the central halo galaxy (see
  Dolag et al. 2013 in preparation for a detailed description).  This
  description of BHs provides a realistic description of the observed
  relation between stellar mass and BH mass, and of the observed
  BH-mass function and luminosity function (see Hirschmann et
  al. 2013, in preparation).

\subsection{Identification of clusters}
The identification of clusters proceeds by running first a FoF
algorithm in the high--resolution regions, which links DM particles
using a linking length of 0.16 times the mean interparticle
separation. The center of each halo is then identified with the
position of the DM particle, belonging to each FoF group, having the
minimum value of the gravitational potential.

Starting from this position, and for each considered redshift, a
spherical overdensity algorithm is employed to find the radius
$R_{\Delta}$ encompassing a mean density of $\Delta$ times the
critical cosmic density at that redshift,
$\rho_c(z)$. 
In the present work, we consider values of the overdensity\footnote{The corresponding radii
  aproximately relate to the virial radius as $(R_{2500}, R_{500},
  R_{200}) \approx (0.2, 0.5, 0.7)\,R_{\rm vir}$
  \citep[e.g.,][]{Ettori2006}.}
$\Delta=2500$, $500$ and $200$.  For the sake of completeness, we
also consider the virial radius which defines a sphere enclosing the
virial density $\Delta_{\rm vir}(z)\rho_c(z)$, predicted by the
spherical collapse model 
($\Delta_{\rm vir}\approx 93$ at $z=0$ and $\approx 151$ at $z=1$ for
our cosmological model).  In total, we end up with about 70 clusters
and groups having $M_{vir}>1\times 10^{14}h^{-1}M_\odot$ at $z=0$.

Throughout this work all the quantities of interest will be evaluated
at the four different characteristic radii.  Therefore, for each
cluster, the hot gas, stellar, and baryonic mass fractions within a
given radius $R_{\Delta}$ are defined, respectively, as 

\begin{equation}
f_{\rm g}(<R_{\Delta}) = \frac{M_{\rm g}(<R_{\Delta})}{M_{\rm tot}(<R_{\Delta})}
\label{eq:fg}
\end{equation}

\begin{equation}
f_{\rm *}(<R_{\Delta}) = \frac{M_{\rm *}(<R_{\Delta})}{M_{\rm tot}(<R_{\Delta})}
\label{eq:fst}
\end{equation}

\begin{equation}
f_{\rm b}(<R_{\Delta}) = \frac{M_{\rm g}(<R_{\Delta})+M_{\rm *}(<R_{\Delta})}{M_{\rm tot}(<R_{\Delta})} \ .
\label{eq:fb}
\end{equation}

\section{Baryon content of clusters}
\label{sec:baryoncontent}

Figures \ref{fig:comparison_physics_fb},
\ref{fig:comparison_physics_fst}, and  \ref{fig:comparison_physics_fg} 
show, respectively, the baryon, stellar, and gas mass fractions of our
simulated clusters as a function of cluster mass $M_{500}$.  
Only clusters with $M_{500}\magcir 3\times 10^{13}\msun$
within our {\tt \nr}, {\tt \w}, and {\tt \agn} runs are shown.  In order
to compare with observational data, we use some representative
observational samples, mainly those from \citet{Lin2003},
\citet{Gonzalez2007}, \citet{Giodini2009}, \citet{Lagana2011}, and
\citet{Zhang2011}.  For all the observational data sets we compare
with, we compile in Table~\ref{tab:fittings} the best fittings
obtained for the baryon, gas, and stellar mass fractions.

Before comparing our results with observational data, let us briefly
describe the main properties of the different observational samples
(for further details, we refer to their corresponding papers).
Knowing how the observational data have been derived is important to
understand not only discrepancies between our simulations and the
observations but also the differences between the observational
results.

\citet{Lagana2011} and \citet{Zhang2011} investigate the baryon mass
content for a subsample of 19 clusters of galaxies extracted from the
X-ray flux-limited sample HIFLUGCS.  For these clusters, the above
authors measure total masses and characteristic radii on the basis of
a rich optical spectroscopic data set, the physical properties of the
intra-cluster medium using \emph{XMM-Newton} and \emph{ROSAT} X-ray
data, and total (galaxy) stellar masses utilizing the DR-7 SDSS
multi-band imaging. Using gas mass measurements from X--ray
observations, \citet{Lagana2011} use a scaling relation between the
gas and the total mass to determine the total cluster mass.  Following
a different approach, \citet{Zhang2011} derive cluster masses from
measurements of the ``harmonic" velocity dispersion as described by
\citet{Biviano2006b}.  In both studies, to obtain the contribution of
galaxies to the stellar mass, they use the optical data for selected
member galaxies within $R_{500}$ to compute their luminosity function
in the i-band, performing an analytical fit using two Schechter
functions. Then, they adopt different mass-to-light ratios for
ellipticals and spirals, taken from \citet{Kauffmann2003} assuming a
\citet{Salpeter1955} IMF, to compute the stellar mass from the optical
luminosity.  In \citet{Lin2003} they use the observed X-ray
mass-temperature relation together with published X-ray emission
weighted mean temperatures, \emph{2MASS} second incremental release
NIR data, and X-ray imaging data to explore trends in the NIR and
X-ray properties of a sample of 27 nearby galaxy clusters.  The total
mass of the clusters in this sample is estimated from an observed
$M_{500}-T_X$ relation. The stellar masses are obtained from the
cluster luminosity function as derived from the magnitudes in the
$K_s$ band. Using a Schechter function they estimate the total cluster
luminosity and, finally, they obtain the total stellar masses, as
contributed by satellite galaxies and BCG, by multiplying the total
luminosity by the average stellar mass-to-light ratio for each cluster
which takes into account the varying spiral galaxy fraction as a
function of the cluster temperature.  In \citet{Giodini2009}, 91
candidate X-ray groups/poor clusters at redshift $0.1 \le z \le 1$ are
selected from the COSMOS 2 deg$^2$ survey, based only on their X--ray
luminosity and extent.  They use X-ray detection, gravitational
lensing signal, optical photometric and spectroscopic data of the
clusters and groups identified in the COSMOS survey. Total cluster
masses are derived from a $L_X-M_{500}$ relation. The stellar mass of
a galaxy is obtained from the conversion of the $K_s$-band luminosity
using an evolving galaxy-type dependent stellar mass-to-$K_s$-band
luminosity ratio.  Since this sample is mostly composed of groups, it
is complemented with the 27 nearby X-ray selected clusters in the
sample by \citet{Lin2003}, where the total and stellar masses are
derived in a consistent manner.  To reduce systematic effects, they
use the scaling relations in \citet{Pratt2009}, based on hydrostatic
mass estimates, to derive the total gas fractions in both samples.
The total sample of 118 groups and clusters with $z \le 1$ spans a
range in M$_\mathrm{500}$ of
$\sim10^{13}$--$10^{15}~\mathrm{M}_{\odot}$.  On the other hand,
\citet{Gonzalez2007} use an optical sample of 24 nearby clusters and
groups for which they obtain drift scan imaging in Gunn $i$ using the
Great Circle Camera on Las Campanas 1 m Swope telescope. This sample
is composed of systems at $0.03 \le z \le 0.13$ that contain a
dominant BCG.  To obtain the total masses and cluster radii they
derive calibrations of the $\sigma-R_{500}$ and $\sigma-M_{500}$
relations using the clusters from \citet{Vikhlinin2006} that also have
published velocity dispersions. They determine the luminosity of the
BCG$+$ICL component by fitting the surface brightness distribution in
each cluster out to a radius of 300 kpc from the BCG. Then, they use
the separate $r^{1/4}$ best-fit profiles of these two components to
build a two-dimensional model image from which they determine the flux
within a given circular aperture. On the other hand, the luminosity of
the cluster galaxies lying within the same aperture is computed by
summing the flux of all galaxies fainter than the BCG and brighter
that $m_I=18$.  Then, they use a relation for the luminosity
dependence of the mass-to-light ratio in the $I$ band to convert from
total luminosity to total stellar mass.  Since they lack measurements
of the mass of hot gas in the ICM for this sample, they fit the
behaviour of the stellar mass fraction with cluster mass and use this
relation to derive total baryon fractions for clusters with published
X-ray gas fractions.  We also compare our results for the gas mass
fractions with a sample of observed groups and clusters at $z \le 0.2$
selected from the X-ray samples of \citet{Vikhlinin2006},
\citet{Arnaud2007} and \citet{Sun2009}.  These authors computed gas
mass fractions at R$_\mathrm{500}$ from hydrostatic mass estimates for
a combined sample containing 41 systems within a range of masses of
[1.5$ \times$10$^{13},$1.1$\times$10$^{15}$] M$_{\odot}$.

Due to the different observational methods used to derive the main
cluster properties, some differences are expected between the baryon
census provided by these samples.  In addition, it is necessary to
point out that, when comparing the stellar mass fractions, only
\citet{Gonzalez2007} take into account the contribution of the ICL
component. In our case we also consider the total (galaxies+ICL)
stellar contribution within clusters.  In the following, after
comparing simulation results to the observed total baryon budget in
clusters and groups, we will dissect the separate contribution of
stars and of hot X--ray emitting gas.

\subsection{Baryon mass fraction}

\begin{figure}
\begin{center}
\scalebox{1.3}{\includegraphics[height=70mm]{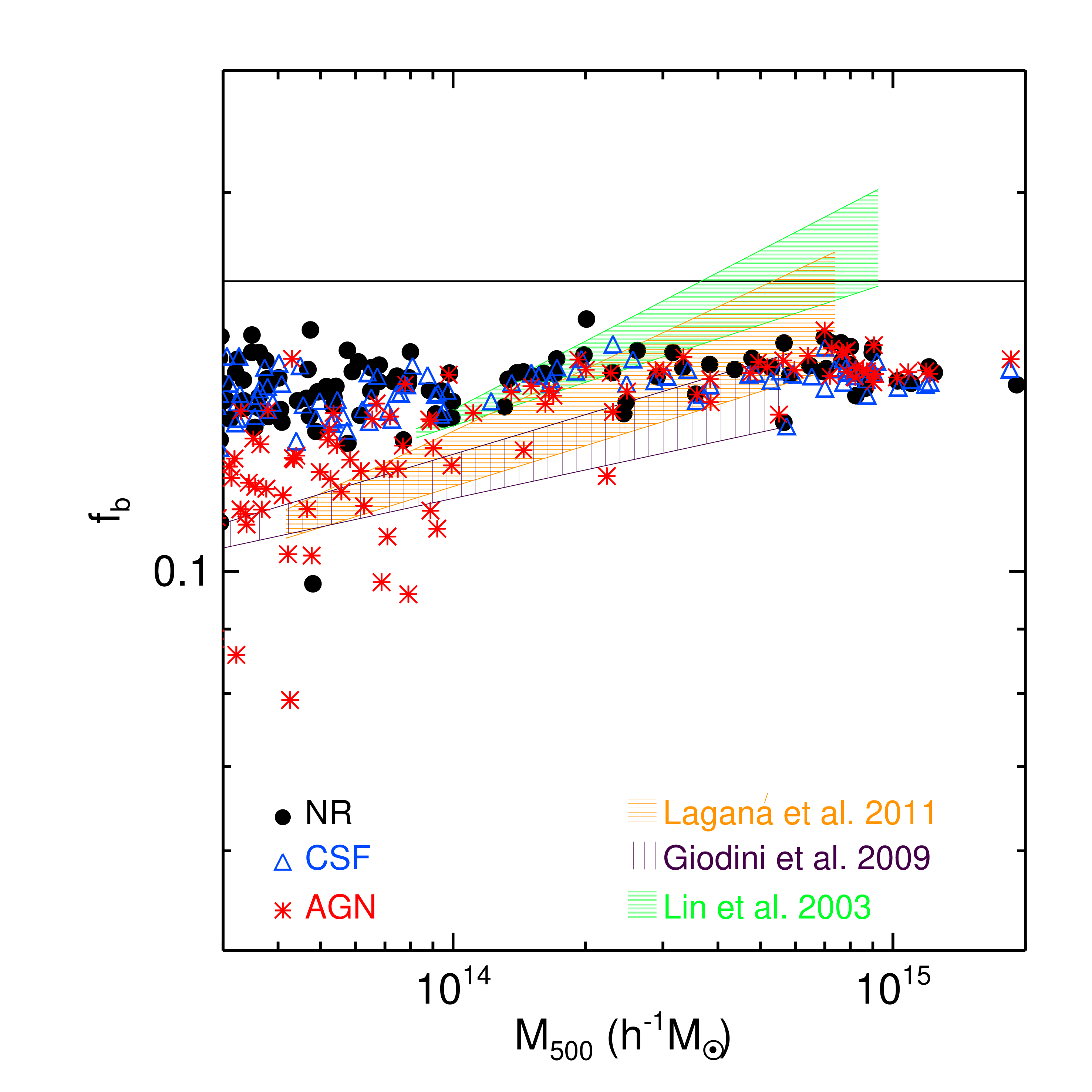}}
\end{center}
\caption{Baryon mass fraction as a function of cluster mass
  $M_{500}$. Results from our {\tt \nr}, {\tt \w}, and {\tt \agn} runs
  are represented by black circles, blue triangles, and red stars,
  respectively.  The observational samples from \citet{Lin2003},
  and \citet{Lagana2011} are shown as shaded
  regions in green and orange, respectively, whereas the sample from 
  \citet{Giodini2009} is represented by a grey-striped area.  
  These regions correspond to the best fits obtained together with
  their corresponding errors (see Table~\ref{tab:fittings}).  The
  horizontal continuos line stands for the cosmic baryon
  fraction assumed in our simulations.}
\label{fig:comparison_physics_fb}
\end{figure} 

As shown in Fig.~\ref{fig:comparison_physics_fb}, in our {\tt \nr}
simulations the baryon mass fractions within $R_{500}$ is nearly
independent of cluster mass and is systematically lower than the
assumed cosmic value by $\mincir 10$ per cent.  This result is
consistent with previous analyses, also based on SPH simulations
\citep[e.g.][]{Eke1998,Ettori2006}, which also found a comparable
underestimate in the cluster baryon fraction. Using a Eulerian AMR
code, \citet{Kravtsov2005} also measured a cluster baryon fraction in
non--radiative simulations below the cosmic value, although in their
case the underestimate was of about 5 per cent within $R_{500}$.
A similar behaviour is also found for our radiative 
{\tt \w} simulations, thus indicating that the processes of star
formation and galactic winds triggered by
SN explosions determine the fraction of baryons to be converted into
stars, without however changing the overall baryon budget within
$R_{500}$.  

As we include the effect of BH feedback in the {\tt \agn} simulations,
there is a significant baryon depletion in poor clusters and groups,
whereas results are nearly the same as for the {\tt \nr} and {\tt \w}
cases for $M_{500}\magcir 2\times 10^{14}\msun$. This result of a
decreasing baryon fraction at low masses is in line with those
presented by \cite{fabjan_etal10}, \cite{Puchwein2010}, and
\cite{mccarthy_etal11}, who also included the effect of BH feedback in
their simulations of galaxy groups and clusters. This effect of baryon
depletion within groups witnesses the efficiency that BH feedback has
in displacing gas outside forming halos. This effect takes mostly
place at relatively high redshift, $z\simeq 2-3$, around the peak of
the BH accretion efficiency. At these epochs, the energy extracted
from BHs increases the gas entropy to levels such to prevent this gas from
being subsequently re-accreted within group--sized halos
\citep[e.g.][]{mccarthy_etal11}.

As for the comparison of simulation results with observations, it is
quite remarkable that a good agreement is only achieved for the
{\tt \agn} model. This result confirms that a feedback mechanisms only
based on SN explosions can not be responsible for the decreasing trend
of the baryon budget within halos of decreasing mass 
\citep[e.g.][]{short_2009}.

\subsection{Stellar mass fraction}

\begin{figure}
\begin{center}
\scalebox{1.3}{\includegraphics[height=70mm]{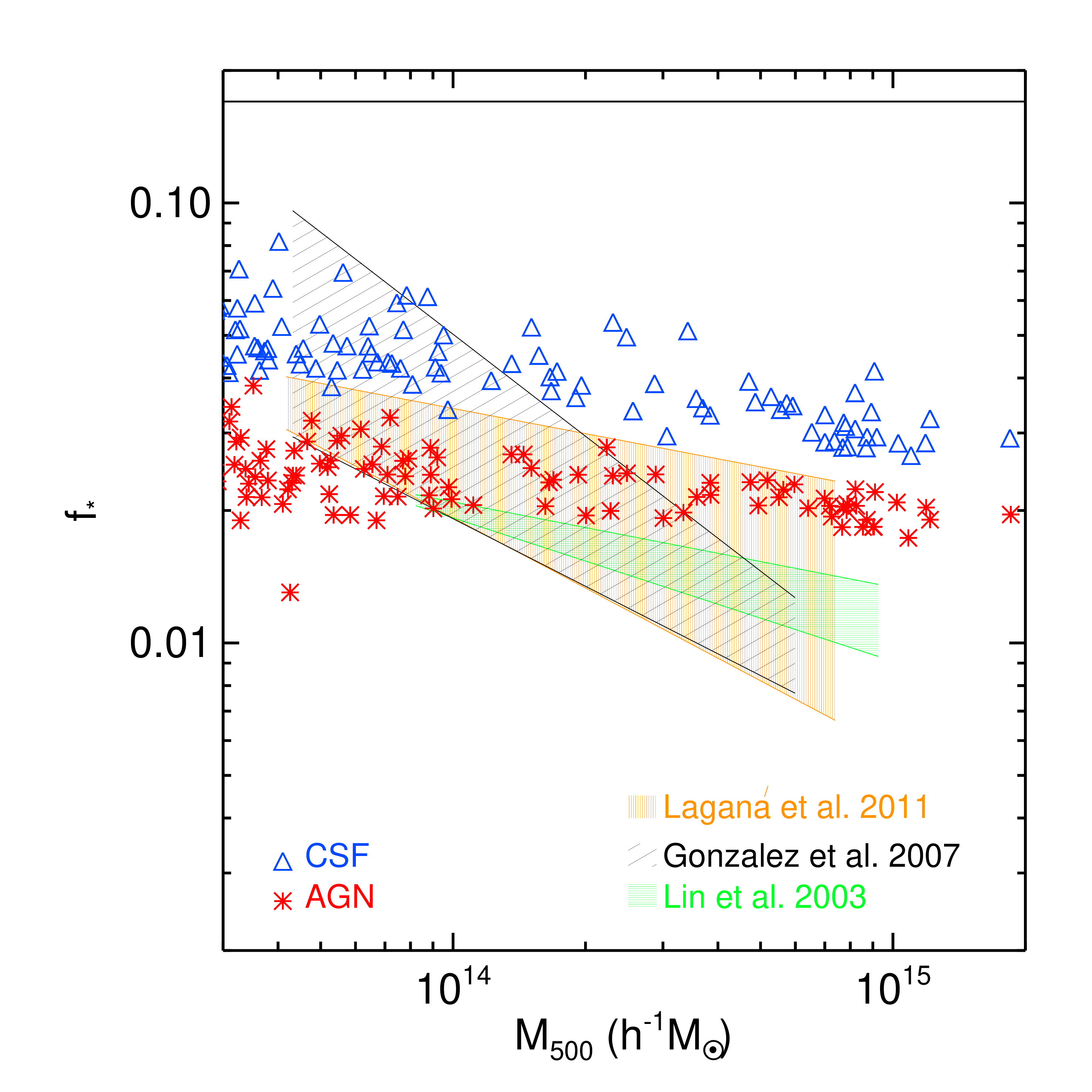}}
\end{center}
\caption{Stellar mass fraction as a function of cluster mass
  $M_{500}$. Results from our radiative simulations, {\tt \w} and {\tt
    \agn}, are represented by blue triangles and red stars, respectively.  The
  observational samples from \citet{Lin2003} and \citet{Lagana2011} 
  are shown as shaded regions in green and orange, respectively, 
  whereas the sample from \citet{Gonzalez2007} is 
  represented by a black-striped area. These regions represent the best fits obtained 
  from each observational sample as compiled in Table~\ref{tab:fittings}.
  The horizontal continuos line stands for the assumed baryon mass
  fraction in our simulations.}
\label{fig:comparison_physics_fst}
\end{figure} 

We show in Fig. \ref{fig:comparison_physics_fst} the stellar mass
fraction  
in our radiative runs. As expected, the effect of including AGN
feedback is that of reducing the stellar content of galaxy systems by
about 30 per cent, nearly independent of cluster mass. As for the
comparison with previous simulation results, we note that the clusters
simulated by \citet{Puchwein2010} with AGN feedback have stellar
fractions, which are larger by about a factor of 2 than 
the stellar fractions found in our {\tt \agn} simulation.  
We can understand this difference by keeping in mind that
the amount of stars formed in simulations depends rather sensitively
on how the SN feedback is included \citep[e.g.,][]{Springel2003b,
  Borgani2006}. In the simulations by \citet{Puchwein2010}, all the
feedback from star formation was injected thermally, without including
kinetic SN feedback as we did in our simulations. As these authors
also noticed (see their Sect. 3.1), 
using kinetic feedback, in addition to thermal feedback,
can significantly reduce the amount of stars formed in their
simulations by a factor of 2, resulting, therefore, in a good
consistency with the stellar mass fractions obtained in our {\tt \agn}
runs.
Common to this kind of simulations is that the BH accretion rate,
  and hence of the amount of feedback energy, is directly derived from
  simulated hydrodynamical quantities by means of a sub--resolution
  accretion model.  Following an alternative approach,
  \cite{Young2011} adopted the hybrid description of
  \cite{short_2009}, which couples a semi-analytic model of galaxy
  formation to a cosmological N-body/SPH simulation, to analyse the
  baryon fractions in clusters of galaxies.  In this approach, the
  energy released to the ICM by SNe and AGN is computed from a
  semi-analytic model and injected into the baryonic component of a
  non-radiative hydrodynamical simulation.  Given that semi-analytic
  models are tuned to reproduce the properties of observed galaxies,
  the main advantage of this approach is that the energetic feedback
  is originated from a realistic population of galaxies.  As a
  potential limitation of this approach, radiative cooling is not
  included in the simulations. As a result, \cite{Young2011} obtain
  stellar fractions in massive clusters that agree better with
  observations than in self-consistent hydrodynamical simulations, but
  they considerably underestimate star formation in groups.

As for the comparison with observational results, we find that our
{\tt \w} simulations produce a too large stellar fraction in massive
galaxy clusters, independent of the observational data set we compare
to. While simulations with AGN feedback give results closer to
observations, the level of agreement is quite sensitive to the
observational result we refer to. For instance, a comparison with the
results by \cite{Gonzalez2007} would imply that in no case simulations
reproduce the steep mass dependence of $f_*$, independently of the
feedback mechanism included \citep[see also][]{andreon10}.
On the other hand, a closer agreement with observations would be
obtained from Fig. \ref{fig:comparison_physics_fst} by referring
instead to the results by \cite{Lagana2011}.  The inclusion of the ICL
component in the analysis by \citet{Gonzalez2007} could explain part
of the difference with respect to \cite{Lagana2011}, although apparently
not all of it \citep[see also][]{Zhang2011}.  Clearly, some caution
must be used when comparing observational and simulated samples, owing
also to the different approaches followed by different authors to
measure stellar mass fraction from data, the dependence of the
inferred stellar mass on the choice of the IMF
\citep[e.g.,][]{Lagana2011,Leauthaud2011}, and systematic
uncertainties in the measurement of the total cluster mass.

\begin{figure}
\centerline{\includegraphics[width=8cm]{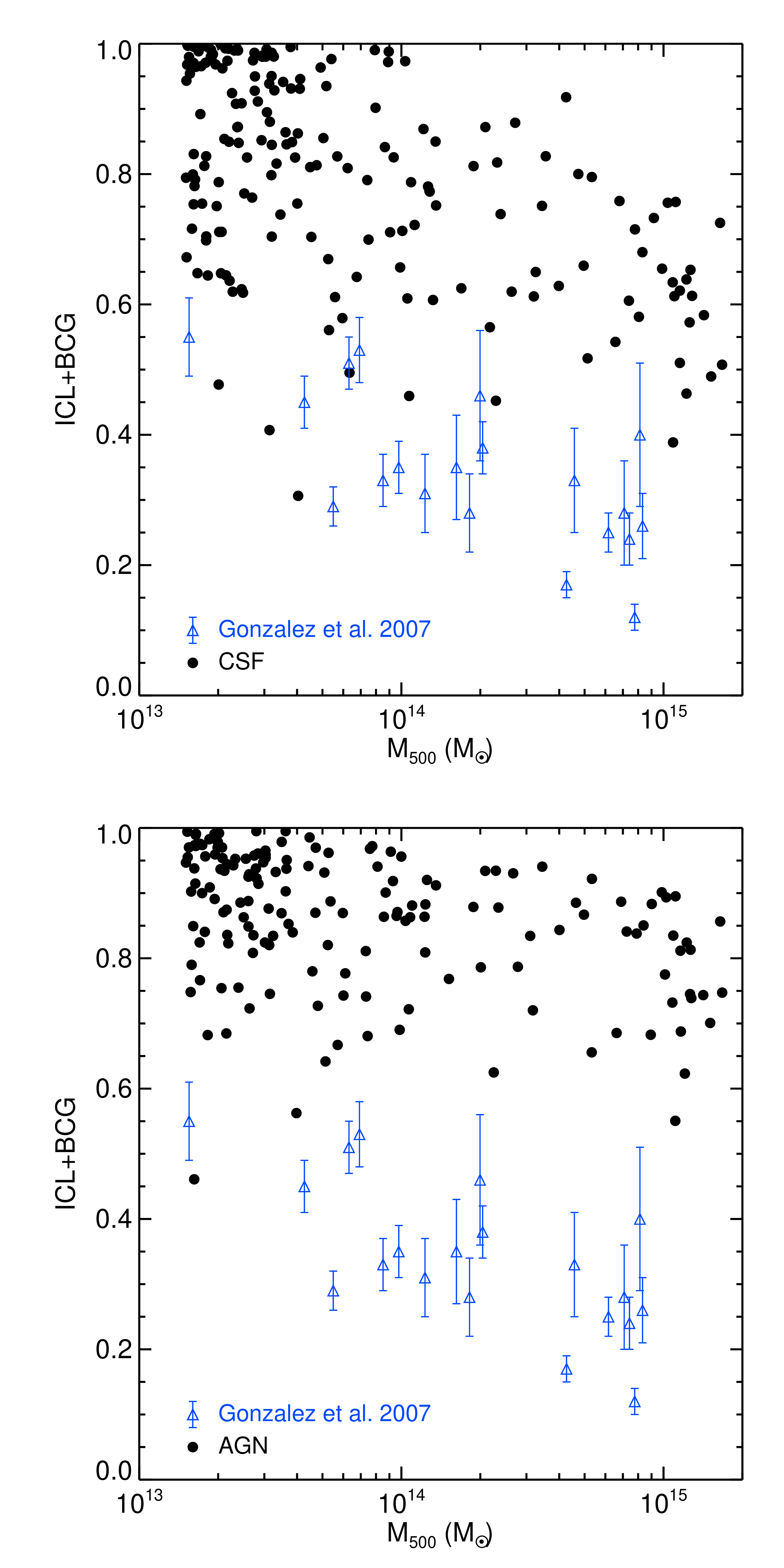}}
\caption{Fraction of the stellar mass found in the BCG+ICL component
as a function of cluster mass $M_{500}$. The
upper panel shows the values obtained from our radiative runs without AGN feedback,
the {\tt \w} runs,  while the lower panel displays the results obtained from our runs including
AGN. We compare our results with the observed BCG+ICL luminosity fractions 
from \citet{Gonzalez2007}.}
\label{fig:icl}
\end{figure}

In order to better understand how much intra--cluster stars contribute
to the total stellar mass budget in our simulations, it is important
to distinguish between the stellar content of the BCGs, of the
satellite galaxies and of the ICL component.  Aside from some
exceptions, the central galaxy in a cluster, which is the closest to
the minimum of the cluster potential well, is typically also the
brightest cluster galaxy. Therefore, for simplicity, we will use the
abbreviation BCG when referring to the central galaxy of a simulated
cluster or group.  Due to its low surface brightness, observations of
the ICL component, which is a smoothly distributed stellar component
typically peaked around the BCG but extended to larger radii, are
difficult, resulting in a significant uncertainty in the current
observational constraints of the amount of ICL present in clusters
\citep[e.g.,][]{Zibetti2005, Gonzalez2007}.

While identifying member galaxies within galaxy clusters in
hydrodynamical simulations is a relatively straightforward task
\citep[e.g.,][]{Onions2012}, distinguishing between stars in the
diffuse stellar component and in the central galaxy is not so
trivial. In order to do so, we use a modified version of the SUBFIND
algorithm \citep{Springel2001, Dolag2008}. In its original version
\citep{Springel2001}, SUBFIND identifies star particles that are
associated with satellite galaxies residing within a cluster--sized
halo. All the star particles not associated to satellite galaxies are
assigned to the central galaxy of the main halo, without
distinguishing between those associated to the actual BCG and those
belonging to the surrounding ICL.   
\cite{Dolag2008} pointed out that
BCG and ICL stars show different phase--space distributions 
and implemented this property in SUBFIND, 
making it able to distinguish among the different stellar components.
For more details about this modified version of SUBFIND we refer 
the reader to the work by \citet{Dolag2009}.

An intrinsic difficulty in properly comparing observations and
simulation results on the amount of ICL is due to the intrinsically
different procedures usually adopted to identify diffuse stars in real
and in simulated data. While observations generally use a criterion
based on surface brightness limit, ICL analysis in simulations is
generally based on identifying star particles that are not
gravitationally bound to galaxies. 
In addition, very faint ICL component can not be detected in observations 
while it is present in simulations.
While we defer to a future paper a
homogeneous comparison between intra-cluster light in observations and
simulations, we show here a comparison between the results by
\cite{Gonzalez2007} on the amount of stars present in BCG and ICL, and
corresponding simulation results. In fact, considering the total
stellar content of BCG and ICL overcomes at least the ambiguity in the
surface brightness limit below which the BCG halo has to be considered
as part of the BCG.  This also avoids choosing among the different ICL
definitions in the literature \citep[e.g.][]{Zibetti2005} allowing,
therefore, a more straightforward comparison between our results and
other simulations and observational studies.  In Fig. \ref{fig:icl} we
plot the fractions of stellar mass found in the BCG+ICL components in
our simulated sample of clusters with $M_{500}\ge 1.5\times10^{13}
M_\odot$.  Results obtained from our radiative simulations without and
with AGN feedback ({\tt \w} and {\tt \agn} runs), are shown in the top
and bottom panels, respectively.  We compare our data with the
observational constraints on the BCG+ICL fraction from
\citet{Gonzalez2007}, shown as blue triangles with error bars.

Although our results in general confirm a decreasing trend with
cluster mass of the fraction of stars contributed by BCG and ICL, they
predict a too large value of this fraction in comparison with the
observational result by \citet{Gonzalez2007}. In the {\tt \w}
simulations we obtain BCG+ICL fractions of roughly $\sim 60$ per cent
for massive clusters, and of about $\sim90$ per cent for groups.
These values are larger than those observed by \citet{Gonzalez2007},
especially for massive clusters.  
However, we have to take into account that there is some uncertainty
in the mass assigned to the BCG and ICL components due to the
analysis method used for separating them \citep[see, for
instance,][]{Puchwein2010}.

In addition, given that the distribution of the BCG+ICL component is
more concentrated than the distribution of the satellite galaxies,
these values are sensitive to the radius inside which they are
measured.

When AGN feedback is included, we find an even larger fraction of
stars in the BCG+ICL component, a result that is consistent with that
presented by \cite{Puchwein2010}.  The reason for this result is that,
although the stellar mass of the BCG+ICL component decreases when
including AGN feedback, the total stellar mass decreases even more
(see Fig.~\ref{fig:comparison_physics_fst}).  A relative increase of
stars in BGC and ICL is the consequence of the combination of two
different effects. On the one hand, the effect of AGN feedback is
mainly that of truncating the star formation of clusters at high
redshift, $z\mincir 3$ \citep[e.g.][]{fabjan_etal10, mccarthy_etal11}.
On the other hand, most of the dynamical origin of the ICL is
associated with the assembly of the BGC
\citep[e.g.][]{Murante2007}. Since mergers continue to take place
after star formation is quenched by AGN feedback, they keep unbounding
stars from galaxies into the diffuse intra--cluster components. Since
this process is not compensated by fresh star formation in the
presence of AGN feedback, the net effect is that of increasing the
fraction of stars that end up in the ICL.  In a future analysis (Cui
et al., in preparation) we will carry out a more detailed comparison
of ICL inventory and properties in simulations, by reproducing in
their analysis the same criteria to identify ICL as in observational
data.

\subsection{Gas mass fraction}

\begin{figure}
\begin{center}
\scalebox{1.3}{\includegraphics[height=70mm]{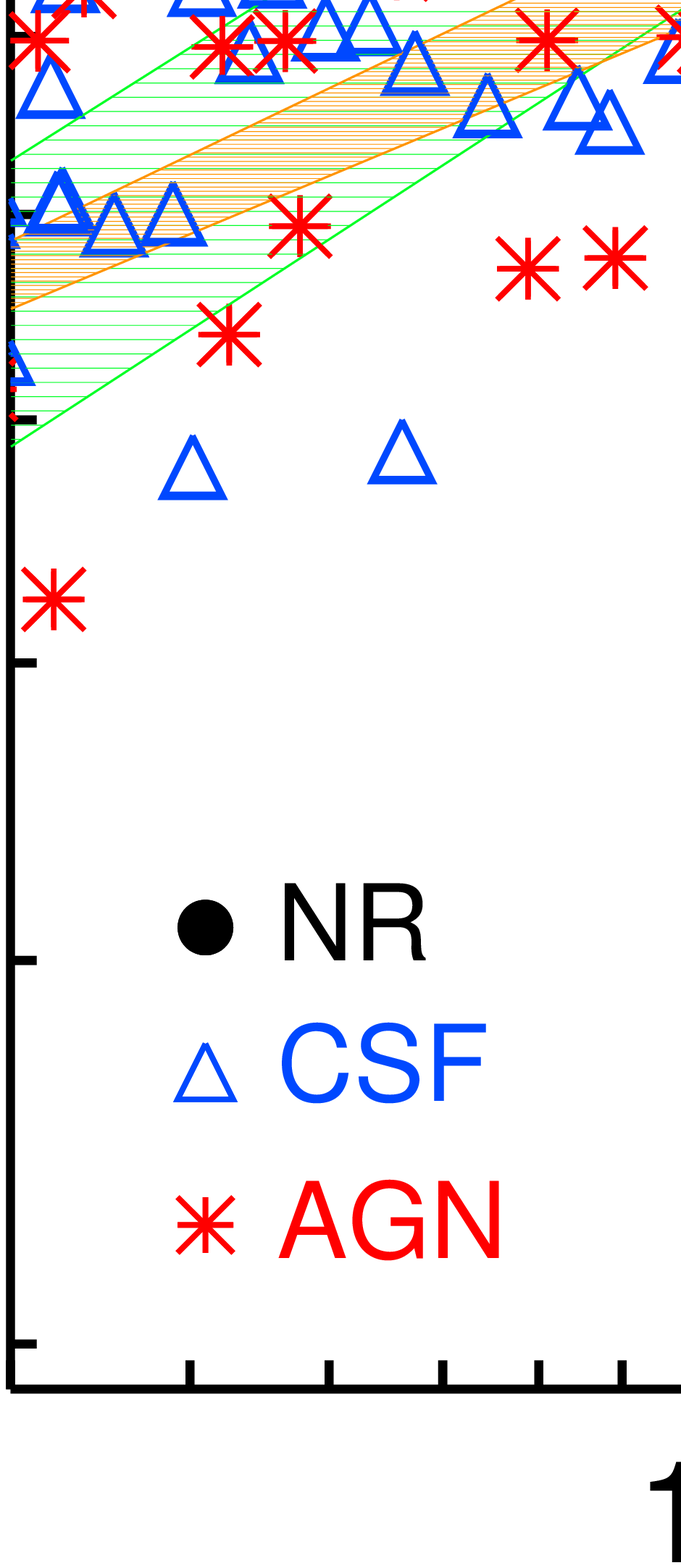}}
\end{center}
\caption{Gas mass fraction as a function of cluster mass
  $M_{500}$. Results from our {\tt \nr}, {\tt \w}, and {\tt \agn} runs
  are represented by black circles, blue triangles and red stars,
  respectively.  We compare our results with two different
  observational samples: a combined sample of 41 clusters and groups
  from \citet{Vikhlinin2006}, \citet{Arnaud2007} and \citet{Sun2009}
  (V06$+$APP07$+$S09), shown as the orange region, and the sample
  obtained from the combination of the data by \citet{Zhang2011} and
  \citet{Sun2009} (Z11$+$S09), shown as the green area (see
  Table~\ref{tab:fittings}).  The horizontal continuos line marks the
  cosmic value of the baryon mass fraction assumed in our
  simulations.}
\label{fig:comparison_physics_fg}
\end{figure} 

As shown in Fig. \ref{fig:comparison_physics_fg}, including
radiative physics in the simulations has the expected effect of
decreasing the gas fraction within $R_{500}$ 
\citep[see also][]{Kravtsov2005, fabjan_etal10, Puchwein2010, mccarthy_etal11,
 Young2011, sembolini_etal12}, by an amount which is more pronounced in poor
clusters and groups. As for the effect of including different feedback
mechanisms, a comparison between our simulations with and without AGN
feedback shows that the two feedback schemes predict rather similar
values of the gas fraction at the mass scale of groups while
simulations including AGN feedback predicts slightly more gas within
large clusters. Clearly, the similar values of $f_g$ in groups do not
imply that feedback does not have any effect on such systems. In fact,
a comparison with Figs. \ref{fig:comparison_physics_fb} and
\ref{fig:comparison_physics_fst} highlights that AGN feedback tends to
remove baryons from the potential wells of galaxy groups. At the same
time, suppression of star formation partially prevents removal of gas
from the hot diffuse phase within $R_{500}$, thereby acting as a compensating effect
such that the resulting gas fraction turns out to be similar for the
two feedback schemes. As for higher--mass halos, AGN feedback becomes
less efficient in removing baryons from halos (see also Fig.
\ref{fig:comparison_physics_fb}), so that suppression of star
formation causes a slightly larger fraction of baryons to remain in
the diffuse phase, so that $f_g$ in this case increases as a result of
a more efficient feedback. This differential effect of AGN feedback in
low-- and high--mass halos is generally quite weak, although it goes
in the direction of better reproducing the observed trend of $f_g$
with halo mass.

From the analysis of Fig. \ref{fig:comparison_physics_fg} we
  conclude that, in general, our results on the values of $f_{g}$,
  especially at the scale of rich clusters, are in better agreement
with the observational results obtained by \cite{Vikhlinin2006},
\cite{Arnaud2007}, \cite{Sun2009}, and \cite{Zhang2011} when AGN
  feedback is included.

\section{Calibration of the baryonic bias}

After having compared simulation results on the different baryonic
components with observational data, in this Section we use our results
to calibrate the different baryonic depletions and to analyse their
dependences on redshift, baryonic physics and cluster radius.

For the sake of comparison with previous works, we define the gas,
stellar and baryon depletion factors (from now on $Y_{\rm g}$, $Y_{\rm
  *}$, and $Y_{\rm b}$, respectively) as the ratios between $f_{\rm
  g}$, $f_{\rm *}$ and $f_{\rm b} = f_{\rm g}+ f_{\rm *}$, and the
cosmic value adopted in the present simulations, $\Omega_{\rm
  b}/\Omega_{\rm m} = 0.167$.  Accordingly, we should measure $Y_{\rm
  b}=1$ within clusters as long as they are fair containers of cosmic
baryons. Any deviation from this value has to be interpreted as due to
the presence of a ``baryonic bias'', whose origin can be due either to
gas dynamical effects at play during the hierarchical assembly of
clusters, or to star formation and feedback effects that causes
sinking or expulsion, respectively, of baryons from the cluster
potential wells. The non-radiative simulations of hot, massive
clusters published by \cite{Eke1998} \citep[see also][]{Crain2007}
give $Y_{b,0}=0.83\pm0.04$ at $R_{2500}$, and are consistent with no
redshift evolution of $Y_b$ for $z<1$. Nevertheless, simulations
including different models of baryonic physics
\citep{Kay2004,Ettori2006, Crain2007, Nagai2007, short_2010} allow for a range of
evolutions. We note, however, that these previous analyses either lack
sufficient statistics of massive systems, which are relevant for
cosmological applications, or the inclusion of an efficient feedback
mechanism, like that provided by AGN, which provides a realistic
description of star formation in the central regions of galaxy
clusters.

The results that we will present in the following are relevant to test
the robustness of the calibration through simulations of the baryon
bias, i.e. of the deviation of the baryon content of clusters from the
cosmic value, that one needs to correct for in the cosmological
application of the gas mass fraction.

\subsection{Radial dependence of the baryonic bias}

\begin{figure*}
\centerline{\includegraphics[width=12cm]{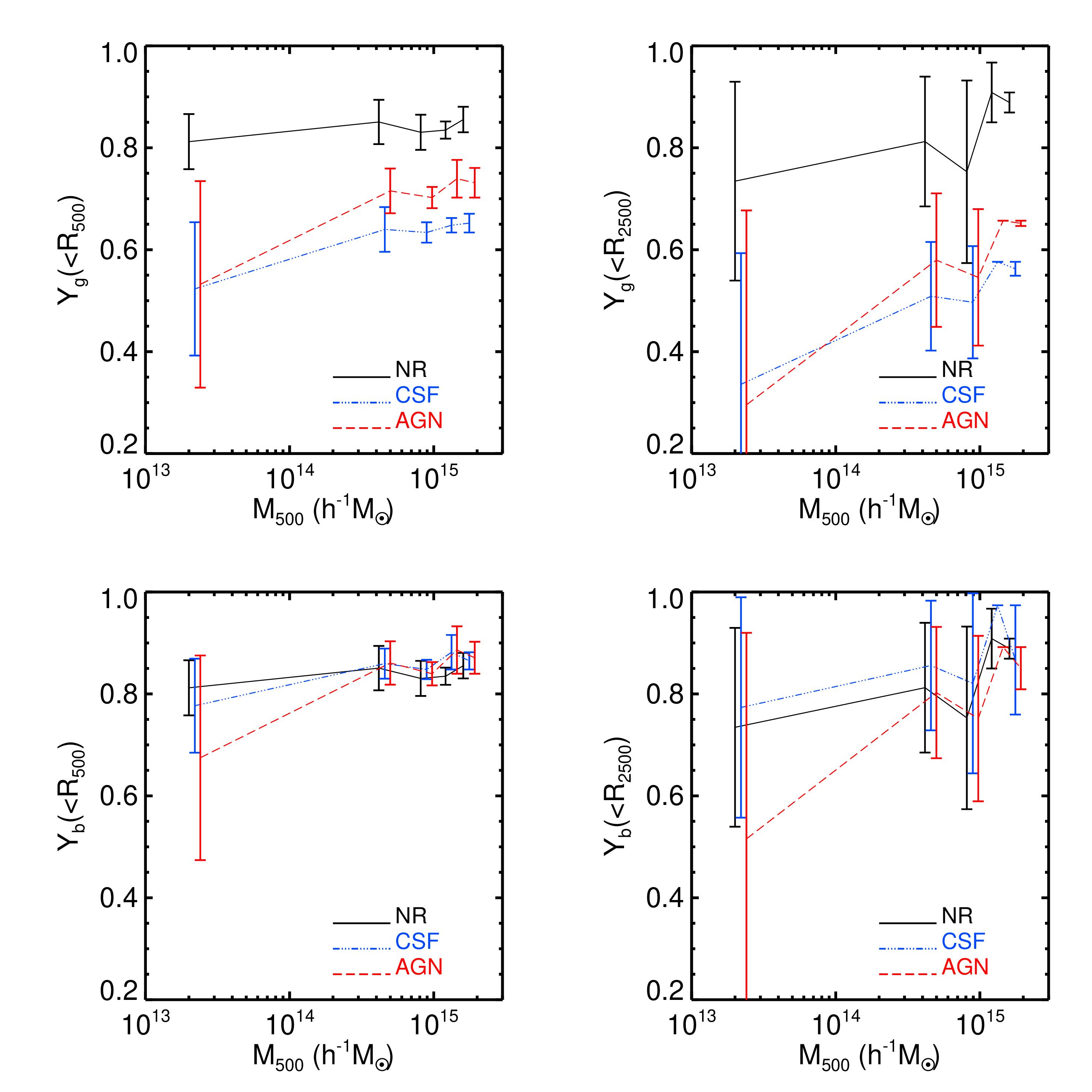}}
\caption{Mass dependence of the gas depletion factor $Y_{\rm g}$
  (upper row), and total baryon depletion factor $Y_{\rm b}$ (lower
  row), computed at $R_{500}$ and $R_{2500}$ (left and right columns,
  respectively). Results are shown for the set of simulated clusters
  with $M_{500}\magcir 1\times 10^{13}\msun$ identified at $z=0$
  within our {\tt \nr}, {\tt \w}, and {\tt \agn} runs.  Clusters are
  binned in five linearly spaced mass bins. Different line types
  represent the mean values obtained within each mass bin for each of
  our simulations, while error bars stand for 1$\sigma$ standard
  deviations computed within the subset of clusters within each mass
  bin. For reasons of clarity, lines corresponding to the different
  physical models have been slightly displaced along the x-axis.}
\label{fig:y_m}
\end{figure*}

We show in Fig.~\ref{fig:y_m} the mass dependence of $Y_{\rm g}$ and
$Y_{\rm b}$ at $z=0$ (up and bottom row, respectively) within the two
characteristic radii $R_{500}$ (left column) and $R_{2500}$ (right
column). As for $R_{500}$, it typically corresponds to the most
external radius out to which detailed X--ray observations, possibly in
combination with Sunyaev--Zeldovich (SZ) observations
\citep[e.g.][]{Planck2011}, allow one to trace the gas content within
clusters, while $R_{2500}$ is the typical radius within which gas
content is traced for distant clusters, when using the evolution of
the gas fraction in clusters as a cosmological probe
\citep[e.g.][]{Allen2008}.
Therefore, for these two radii, we show the mean values of the gas and baryon 
depletion factors obtained for our
sample of simulated clusters binned in five linearly equi-spaced bins
in $M_{500}$.  All clusters with $M_{500}\magcir 1\times 10^{13}\msun$
have been considered.  The mean values within each mass bin are shown
along with error bars representing one standard deviation within the
corresponding mass interval.

The left panel of this figure summarizes the simulations results shown
in Figs.  \ref{fig:comparison_physics_fb}  and  \ref{fig:comparison_physics_fg}. 
Using the mass binning, it is now
more clear that the depletion in baryon content within $R_{500}$ is
more pronounced and with a stronger mass dependence for the
simulations including AGN feedback, at least for low--mass systems. 
As already discussed, the larger baryon depletion in clusters simulated
with AGN feedback is the result of the efficiency of this feedback
mechanism in removing baryons from the potential wells of forming
groups at redshift $z\sim 2$--3, around the peak of gas accretion
onto SMBHs. On the other hand, for masses $M_{500}\magcir 2\times
10^{14} \msun$ we find that the baryon fraction within $R_{500}$
underestimates the cosmic value by about 15 per cent, nearly
independent of mass and of the physics included in the
simulations. The r.m.s. dispersion around this values is of about 3
per cent for the {\tt \nr} and {\tt \w} simulations, which increases
to about 5 per cent for the {\tt \agn} simulations. This result for
$Y_b$ is different from the behaviour of gas depletion, which shows no
flattening for high--mass systems. Furthermore, values of $Y_g$ for
the AGN simulations are systematically larger than for the radiative
simulations including only SN feedback, as a result of the suppressed
star formation in the former case.

These results suggest that a mass--independent correction can be
calibrated from simulations to infer the cosmic baryon fraction from
the corresponding quantity derived for massive clusters, a correction
that is likely independent of the uncertain knowledge of the physical
processes at play in the ICM. However, these results also highlight
that accurately recovering the baryon fraction from gas mass
measurements involves accurately accounting for a correction
associated to stellar mass, which generally depends on cluster total
mass.

Results on gas and baryon depletions are somewhat different within
$R_{2500}$ (right column of Fig. \ref{fig:y_m}). In this case, both
$Y_g$ and $Y_b$ show as steady increase with cluster mass, with no
evidence for a flattening at high masses, and a larger intrinsic scatter in
their values. Furthermore, a small but sizeable dependence of $Y_b$ on
the physics included in the simulations exists even for the highest
mass systems. Quite interestingly, the largest values of $Y_b$ are
obtained for the {\tt \w} simulations, as a result of the strong
overcooling, not efficiently counteracted by the SN feedback, which
causes a large amount of baryons to condense in the central halo
regions. On the other hand, the smallest $Y_b$ value is obtained in
the presence of AGN feedback, which is quite efficient in displacing
gas from central regions even for the most massive clusters. In
general, these results indicate that the baryon depletion within
$R_{2500}$ is more sensitive to the detailed description of the
feedback process which regulates the cooling--heating cycle. As a
result, care must be taken in the use of simulations to exactly
calibrate the correction for baryon depletion to observations
providing gas and baryon fractions at such smaller cluster--centric
radii.

\begin{figure*}
\begin{center}
\scalebox{1.4}{\includegraphics[width=5cm]{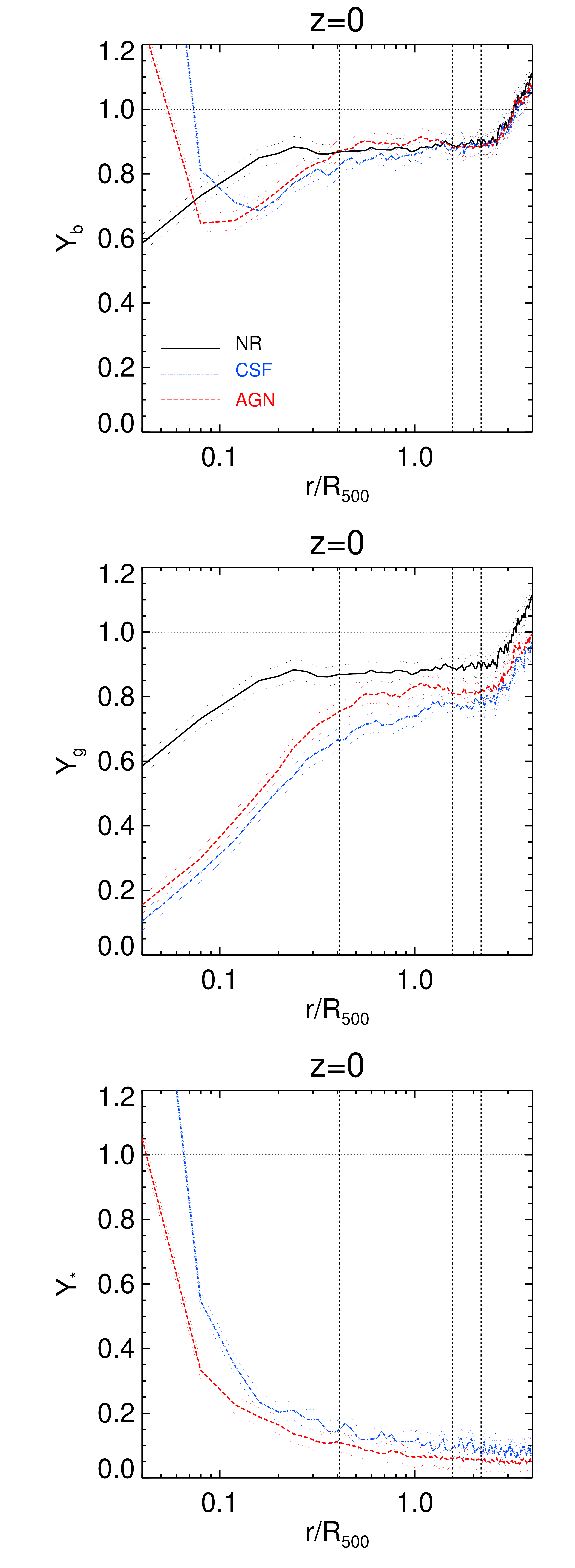}}
\scalebox{1.4}{\includegraphics[width=5cm]{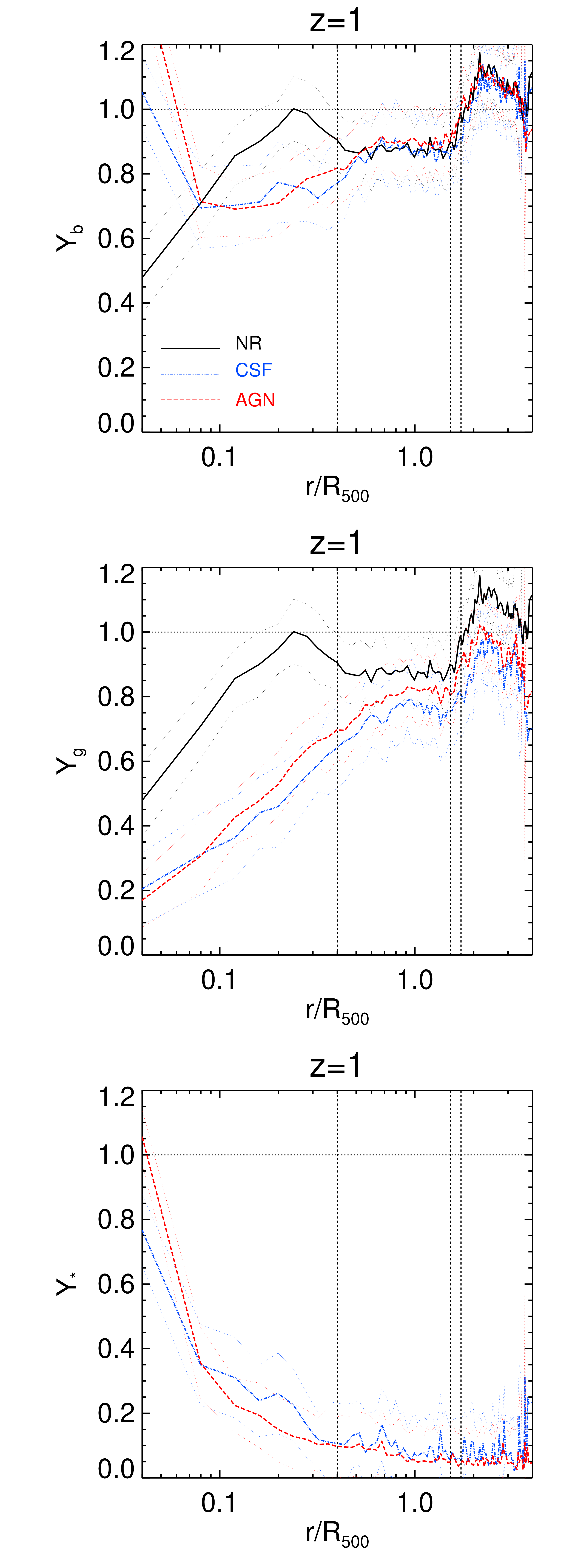}}
\end{center}
\caption{The mean radial profiles out to $4\times R_{500}$ of the
  baryonic, gas and stellar depletions (from top to bottom panels) at
  $z=0$ (left column) and $z=1$ (right column)
  for the subsample of massive clusters with $M_{500}>2\times
  10^{14}\msun$.
  In each panel, mean profiles for the {\tt \nr}, {\tt \w} and {\tt
  \agn} simulated clusters are shown with the continuous black,
  dashed-dot blue, and long-dashed red lines, respectively.  For each
  model, dotted lines around the mean profiles indicate the region
  corresponding to $1\sigma$ standard deviation around the mean. Within
  each panel and from left to right, dotted vertical lines indicate the
  position of the mean value of $R_{2500}$, $R_{200}$ and $R_{vir}$ (in
  units of $R_{500}$) for the sample of clusters for which the profiles
  are computed.}
\label{fig:radial_depletions}
\end{figure*} 

As shown in Fig.~\ref{fig:y_m}, the population of the most massive
clusters is characterized, especially within $R_{500}$,
both by a remarkably small intrinsic scatter
in their baryon budget, and by a stability against the different
physical descriptions of the ICM. As such these massive clusters are
those which can be more reliably calibrated for cosmological
applications of the cosmic baryon fraction test. Therefore, in the
following we will restrict the analysis of these radial distributions
only to clusters with $M_{500}> 2\times 10^{14}\msun$.  Within our
simulations we identify about 40 of such objects at $z=0$, a number
that reduces to 10 at $z=1$.

The effect of the different physical models considered in our
simulations on the distribution of baryons can be better understood
from the analysis of the radial distribution of the different baryonic
components within clusters. 
Figure~\ref{fig:radial_depletions} shows the mean radial distribution
out to $4\times R_{500}$ of the baryonic, gas and stellar depletions at
$z=0$ (left column) and $z=1$ (right column) for our subsample of
massive clusters within each of the physical schemes adopted in our
simulations.

Regardless of the baryonic processes included in our re-simulations,
the baryonic depletion at $z=0$ for radii $r/R_{500} \ge 0.4$
approaches a value of $\sim 85$ per cent of the cosmic value, showing
similar values at $z=1$ but with larger scatter.  This baryonic
depletion starts to converge to the expected value at $r \sim 3\times
R_{500}$, consistent with results found in previous simulations
\citep[e.g.,][]{Eke1998, Ettori2006}.  In these outer regions of
clusters, the gas mass dominates the baryon budget.  In general, the
gas depletion increases from inner to outer regions and shows slightly
higher values at low redshifts.  On the contrary, the stellar
depletion decreases when moving towards more external regions and,
therefore, if we move towards more internal radii ($r/R_{500} \le
0.1$) the stellar mass clearly dominates the baryon content in the
radiative runs.  In these central regions, the non-radiative
simulations produce lower values of the baryonic depletion than the
radiative runs which are, indeed, characterised by a steep inner
slope.  When cooling and star formation are included, the gas can cool
and form stars and, therefore, it sinks deep into the potential wells
of the clusters.  If AGN feedback is also added, its main effect is
that of heating the surrounding gas producing, therefore, smaller
baryon mass fractions in the cluster center.  In general, the
different baryonic depletions obtained in the inner regions of
clusters from the {\tt \w} and {\tt \agn} simulations are comparable
with each other.  
However, we note that while the baryon and stellar depletions at
  $z=1$ are higher in the {\tt \w} simulations, this is inverted at
  $z=0$. The reason for this lies in the stronger feedback associated
  to galactic winds in the {\tt \w} model. In fact, the effect of AGN
  feedback at $z=1$ is still sub-dominant. As a result, at this
  redshift the higher wind efficiency of {\tt \w} with respect to {\tt
    \agn} model turns into  a reduced effect of cooling, which
  manifests itself both in the total amount of baryons sinking in the
  cluster potential wells and in the amount of stars produced. It is
  only at $z<1$ that AGN feedback prevails, thereby reverting these trends.

\begin{figure}
\centerline{\includegraphics[width=9.5cm]{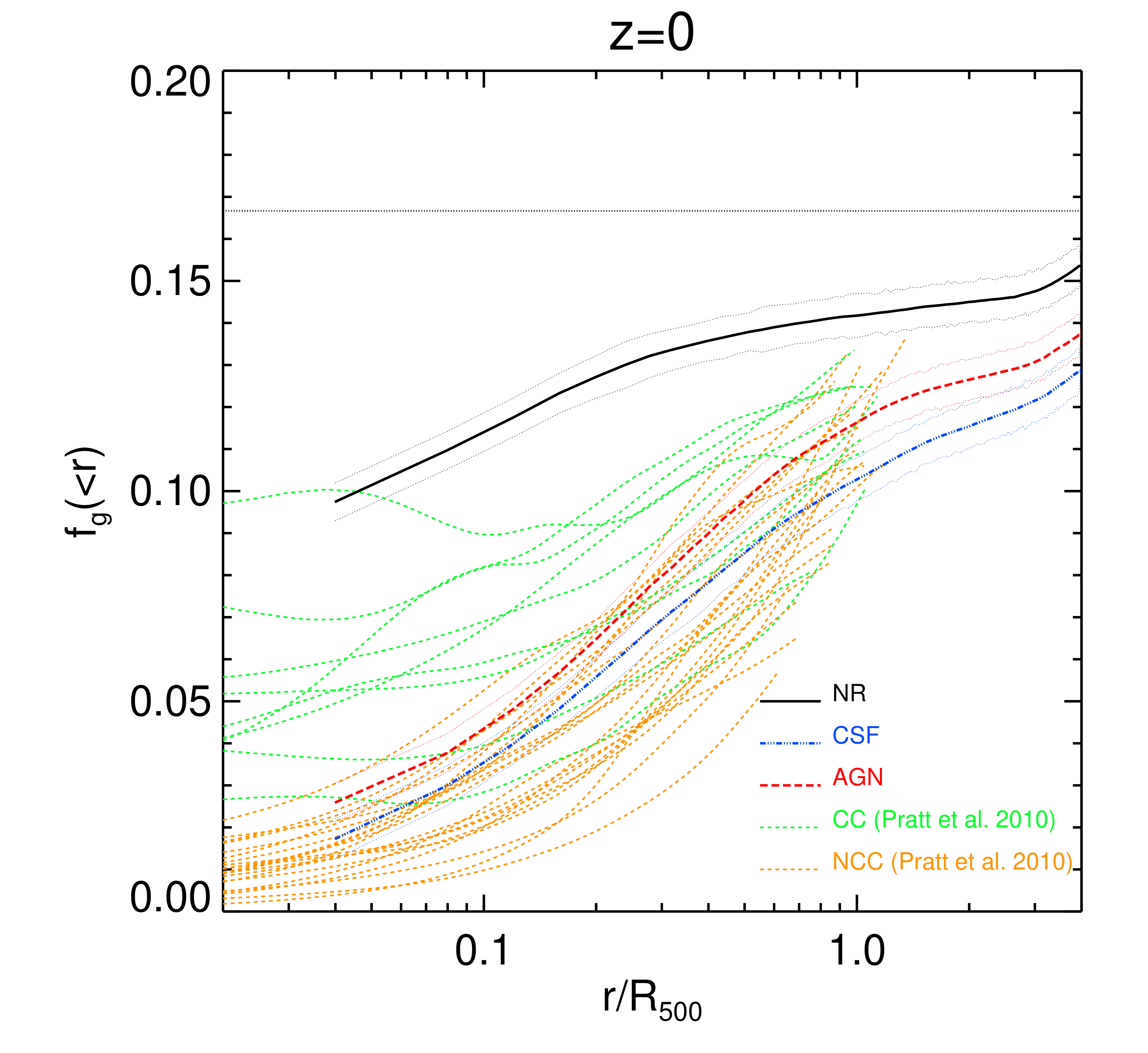}}
\caption{Comparison at $z=0$ between the cumulative gas mass fraction
  profiles, $f_g(<r)$, for the clusters within each of the physical
  schemes adopted in our simulations and the observational data from
  \citet{Pratt_2010}.  The simulated radial profiles are computed out
  to $4\times R_{500}$ for the subsample of massive clusters with
  $M_{500}>2\times 10^{14}\msun$.  Profiles for the {\tt \nr}, {\tt
    \w} and {\tt \agn} simulations are shown with the continuous
  black, dashed-dot blue, and long-dashed red lines, respectively.
  For each model, dotted lines around the mean profiles indicate the
  region corresponding to $1\sigma$ standard deviation around the
  mean.  Dashed lines in orange and green stand for the sample of NCC
  and CC systems from \citet{Pratt_2010}.  The horizontal dotted line
  marks the cosmic value of the baryon mass fraction assumed in our
  simulations.}
\label{fig:fg_extra}
\end{figure}

In Fig.~\ref{fig:fg_extra} we show the comparison between the
cumulative gas mass fraction profile, $f_g(<r)$, in our simulations
and from the observational results by \citet{Pratt_2010}. The
cumulative gas profiles within each of our physical models have been
computed at $z=0$ out to $4\times R_{500}$ for our subsample of
massive clusters, which lay in the temperature range $2-10$ keV.
\citet{Pratt_2010} examine the radial entropy and gas distributions of 31 
nearby galaxy clusters from the Representative \xmm\ Cluster Structure Survey
\citep[REXCESS,][]{Bohringer_2007}. This sample, which includes
clusters with temperature in the range 
2--9 keV, has been selected in X-ray luminosity only, with no bias towards any 
particular morphological type. According to their central densities,  clusters in 
this sample have been classified in cool core systems (CC) and in morphologically 
disturbed or non-cool core  (NCC) systems\footnote{More especifically, this sample 
of 31 clusters contains 10 CC and 12 NCC systems. The rest of systems, which have 
not been morphologically classified, are shown in Fig.~\ref{fig:fg_extra} as NCC clusters.}.

As we can infer from Fig.~\ref{fig:fg_extra}, our radiative simulations are quite
successful in reproducing the observed gas profiles for the NCC population out to the 
limit of the observations.  However, as already reported by other authors 
\citep[see, for instance,][]{Young2011},  radiative simulations fail in matching the profiles 
associated with the CC clusters, which show flatter gas mass fraction profiles in 
their inner regions.

\subsection{Redshift evolution of the baryonic bias}

Besides assessing the stability of the baryon and gas depletions at
$z=0$, measuring the evolution of such quantities is also required for
measurements of the gas fraction over a large redshift baseline to be
used to recover the redshift--distance relation.  A comparison of the
profiles of baryon and gas depletion computed at $z=0$ and $z=1$ (see
Fig. \ref{fig:radial_depletions}) shows that outside the central
regions this evolution is generally rather mild.

\begin{figure*}
\centerline{\includegraphics[width=16cm]{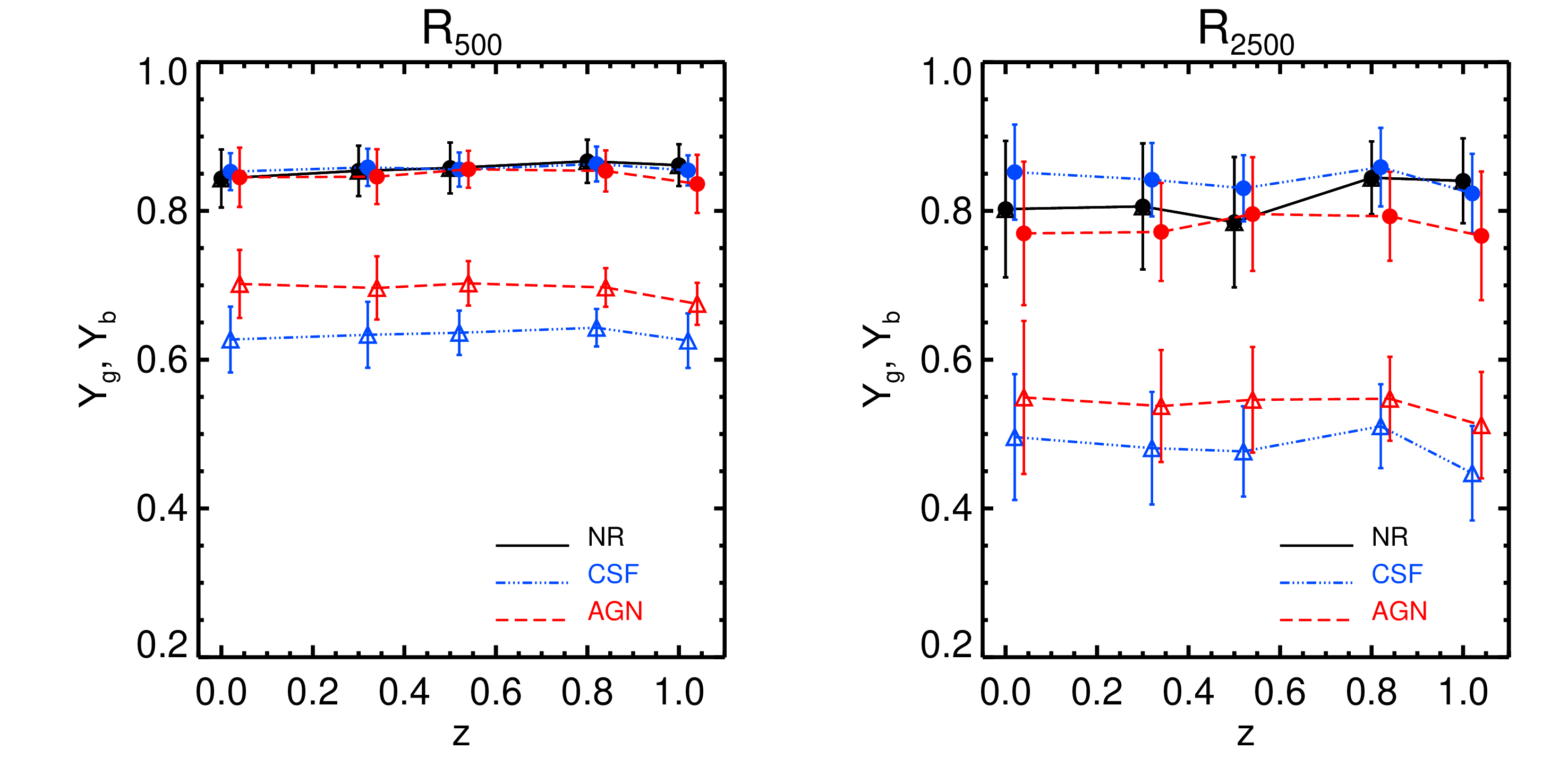}}
\caption{The redshift dependence of the mean values of gas depletion
  $Y_{g}$ (triangles) and baryon depletion $Y_b$ (circles), computed
  at $R_{500}$ (left panel) and $R_{2500}$ (right panel), for all
  clusters that at each redshift have mass $M_{500}>2\times
  10^{14}\msun$. In each panel, continuous black,
  dashed-dot blue,  and long-dashed red lines stand for
  the {\tt \nr}, {\tt \w}, and {\tt \agn} simulation sets,
  respectively.  
  These lines have been slightly displaced along the x-axis to avoid overlapping
  among them.
  Error bars represent $1\sigma$ intrinsic scatter
  computed over all simulated clusters.}
\label{fig:y_z}
\end{figure*}

To quantify this evolution, we compute the values of the depletion
factors at different redshifts, $z=0$, $0.3$, $0.7$, $0.8$ and $1$. At
each redshift, the analysis is done only on those clusters having mass
$M_{500}> 2\times 10^{14}\msun$. We show the redshift evolution of
these quantities at $R_{500}$ and $R_{2500}$ in the left and right
panels of Fig. \ref{fig:y_z}, respectively, along with their
respective intrinsic scatters, given by the error bars. 
A compilation of these values, for the different redshifts and 
simulation sets, are also reported in Table \ref{tab:fgas}.

In order to parameterize a possible evolution of the values of the
depletion factors, we use the expression
\begin{equation}
Y_i(z)=Y_{0,i}(1+\alpha_{Y_i}z)\,,
\label{eq:fit} 
\end{equation}
where the subscript $i$ is equal to $b$ or $g$ when referring to the
total baryon or gas content, respectively.
The values of the parameters $Y_{0,i}$ and $\alpha_{Y_i}$ are computed
through a $\chi^2$ minimization procedure, with the weights of the
data points reported in Table \ref{tab:fgas} provided by their
corresponding intrinsic scatter.

In Table \ref{tab:ajustes} we report the best--fitting values of these
parameters obtained for each simulation set, within different radii of
interest, from $R_{2500}$ out to the virial radius $R_{vir}$.

The results displayed in Fig. \ref{fig:y_z} show that, independently
of the considered radius or physics, the baryon depletion factor does
not evolve significantly with redshift, at least since $z=1$.  Within
$R_{2500}$ (right panel) the dissipative action of radiative cooling
in the {\tt \w} runs slightly increases the average value of $Y_{b}$
with respect to the {\tt \agn} simulations, bringing it very close
or even above to
that of the non--radiative simulations, with $Y_{b}\simeq 0.85$,
constant across the considered redshift range. On the other hand, the
presence of AGN feedback is effective in preventing gas from accreting
onto the central regions, thus decreasing the baryons fraction to
$Y_{b}\simeq 0.80$, also independent of redshift.

As for results at $R_{500}$, we find a smaller scatter and much better
agreement among the different physical models, thus highlighting that
the different physical descriptions of the ICM have a negligible
impact on the total amount of baryons at such larger cluster--centric
radii. At such radii, we find $Y_{b}\simeq 0.85$ virtually independent
of redshift, with some departure for the {\tt \agn} simulations at
$z=1$, probably due to the limited statistics of massive clusters at
the highest considered redshift. Therefore, a sizeable decrease in the
baryon fraction when moving inwards to $R_{2500}$ is detected when
including the more realistic feedback scheme based on the effect of
AGN.

As for the gas mass fraction, the inclusion of radiative physics
decreases its value with respect to the non--radiative simulations,
both at $R_{2500}$ and at $R_{500}$. As expected, this decrease is
more pronounced at smaller radii and for the simulations only
including the effect of SN feedback.  As for the {\tt \agn}
simulations, we find $Y_g\simeq 0.5$--0.6 within $R_{2500}$, quite
independent of redshift, with a significant scatter,
$\sigma_{Y_g}\simeq 0.1$, over the whole range of redshift. This value
increases to $Y_g\simeq 0.6$--0.7 within $R_{500}$, also nearly
constant in redshift, but with a reduced intrinsic scatter of
$\sigma_{Y_g}\simeq 0.05$.  The behaviour obtained for the {\tt \w}
simulations is pretty similar but with lower values for the gas
fraction: within $R_{2500}$, $Y_{g}\simeq 0.5$, whereas it is
$Y_g\simeq 0.6$--0.7 at $R_{500}$. Quite remarkably, in all cases
such values are nearly independent of redshift.

In general, our results for the {\tt \nr} case are in agreement with
those from non-radiative simulations presented by \citet{Eke1998} and
\citet{Ettori2006}, both based on SPH simulations, while they are
slightly, but systematically, lower by about 5 per cent than those
obtained by \citet{Kravtsov2005} from AMR simulations.  Although this
difference is quite small, it is still comparable to, or larger than,
the difference induced by the presence of different physical processes
in simulations. Although it remains to be seen whether such a
difference between predictions of SPH and AMR codes persists when
including radiative physics, its presence warns on the need of
understanding in detail the performances of different hydrodynamical
methods in the calibration of the gas mass fraction test through
simulations. In general, our results on non--radiative simulations
including only SN feedback are in line with those presented by other
authors
\citep[e.g.][]{Muanwong2002,Kravtsov2005,Ettori2006,sembolini_etal12}. However,
while the comparison between results from different non--radiative
simulations is relatively straightforward, when extra-physics is
included the results on the distribution of the different baryonic
components are sensitive not only to the nature of the feedback
sources included (i.e. SNe vs. AGN), but also to the details of the
numerical implementation (i.e. thermal vs. kinetic feedback,
dependence of cooling rates on local gas metallicity, numerical
resolution). These aspects have to be taken into account when
performing such a comparison between results from different authors.

In principle, the results of our analysis can be used to set priors on
the parameters which determine the amount of gas depletion within a
given aperture radius and its redshift evolution, when deriving
cosmological parameters from observations of the baryon fraction in
massive clusters. Overall, the results presented in Fig. \ref{fig:y_z} 
(see also Table \ref{tab:ajustes}) allow us to set a
rather strong prior on the parameter describing the evolution of $Y_b$
(see Eq. \ref{eq:fit}), with $-0.02 \leq \alpha_{Y_b} \leq 0.07$, for a
conservative range of variation holding both at $R_{500}$ and
$R_{2500}$, accounting for the difference between different physical
models and for the uncertainties in the estimate of the mean
associated to the measured intrinsic scatter. As for the
normalization, a conservative allowed range of variation can be taken
to be $Y_{0,b}=0.81\pm 0.06$ at $R_{2500}$, 
with a slightly larger normalization and narrower uncertainty
($0.85\pm0.03$) at $R_{500}$.

In order to go one step further, we analyse the dependences 
of the gas and baryonic  depletions as a function of
any combination of the parameters $\{z, M, \Delta\}$ using the  following expression:

\begin{equation}
Y_i(z)=Y_{0,i}(1+\alpha_{Y_i}z) (M/M_0)^{\beta_{Y_i}}(\Delta/\Delta_0)^{\gamma_{Y_i}} \,,
\label{eq:fit_extra} 
\end{equation}
where $M_0=5.0\times 10^{14} \msun$, $\Delta_0=500$, and the subscript
$i$ stands for $b$ or $g$ when referring to the baryon or gas
depletions, respectively.  
To obtain the values of the different parameters
[$Y_{0,i}$, $\alpha_{Y_i}$, $\beta_{Y_i}$, $\gamma_{Y_i}$],
the least-squares fit has been performed with the IDL routine MPFIT 
\citep{Markwardt_2008}.
For each fit we
have also considered the case in which $\alpha_{Y_i}$ is fixed to 0,
i.e., no evolution.  In Table \ref{tab:ajustes_extra} we report the
best--fitting values of these parameters obtained for each simulation
set. In order to obtain these dependences we have considered the
baryon content of our sample of massive clusters for $\Delta=2500,
500, 200$.

In general, we note that no significant dependences upon $\{z, M,
\Delta\}$ are detected.  All the fitted parameters for $[\alpha_{Y_i},
\beta_{Y_i}]$ have values $\leq 0.07$, whereas $\gamma_{Y_i}$ is in
the range $[0.0, -0.04]$ for $Y_b$ and $[0.0, -0.13]$ for $Y_{g}$.  

\section{Summary and discussion}

In the present study, we have analysed a set of cosmological
hydrodynamical simulations of galaxy clusters paying special attention
to the effects that different implementations of the baryonic physics
have on the baryon content of these systems. Using the newest version
of the parallel Tree--PM SPH code {\tt GADGET-3} \citep{springel05},
we carried out re-simulations of 29 Lagrangian regions extracted
around as many galaxy clusters identified within a low--resolution
N-body parent simulation. These cluster re-simulations have been
performed using different prescriptions for the baryonic physics:
without including any radiative processes ({\tt \nr} runs), including
the effect of cooling, star formation, SN feedback ({\tt \w} runs),
and including also an additional contribution from AGN feedback ({\tt
  \agn} runs).

The final sample of objects obtained within each one of these set of
re-simulations consists in $\simeq 160$ galaxy clusters and groups
with $M_{vir}\ge3\times 10^{13}\msun$ at $z=0$.  Using these three
sets of simulated galaxy clusters, we have analysed how the different
physical conditions within them affect their baryon, gas and stellar
mass fractions and how these results compare with observations.  We
have also examined which are the implications of our results on the
systematics that affect
the constraints on the cosmological
parameters obtained through the evolution of the cluster baryon mass
fraction. In order to do so we have calibrated the different baryonic
depletion factors and we have analysed their dependences on redshift,
baryon physics, and cluster radius.

Our main results can be summarised as follows.
\begin{itemize}

\item In the {\tt \nr} simulations the baryon mass fractions within
  $R_{500}$ appear flat as a function of the total mass and differ by
  less than $\sim10$ per cent from the assumed cosmic baryon fraction.
  This result  is
  consistent with previous non-radiative
  simulations \citep[e.g.,][]{Eke1998, Kravtsov2005,Ettori2006}.
  Whereas the {\tt \w} simulations present a similar behaviour, when
  AGN feedback is included there is a significant baryon depletion in
  poor clusters and groups, whereas the cosmic value holds only for
  the most massive clusters.  This result, which is in agreement with
  the trend displayed by the observational samples from
  \cite{Lin2003}, \cite{Giodini2009}, and \cite{Lagana2011},
  highlights the efficiency of the AGN heating in displacing large
  amounts of gas outside the potential wells of small clusters and
  groups.
\item The stellar mass fractions obtained in our radiative runs, both
  {\tt \w} and {\tt \agn}, decreases smoothly with increasing cluster
  mass and shows a flattening in the low-mass end $(\le 10^{14}
  M_{\sun})$ of our sample.  When comparing with observational data,
  the obtained stellar mass fractions in our {\tt \w} runs is quite
  large, especially for massive clusters \citep[e.g.][]{Lin2003,
    Gonzalez2007, Lagana2011}.  When AGN feedback is included, the
  stellar mass fractions within $R_{500}$ are lowered by about one
  third, thus alleviating the tension with observations, especially at
  the scale of intermediate-mass clusters.  However, the level of
  agreement with observations depends on the observational sample we
  compare with.  Whereas none of our runs is able to reproduce the
  observed strong trend of the stellar mass fraction with cluster mass
  reported by \cite{Gonzalez2007}, our results for the {\agn} runs are
  in closer agreement with other observational samples
  \citep[e.g.,][]{Lagana2011}.
\item When analysing the different stellar components separately, we
  find that the fraction of stars within $R_{500}$ found in the
  BCG+ICL components is, in the {\tt \w} runs, of about $60$ per cent
  for massive clusters, and of about $90$ per cent for groups.
  Paradoxically, when AGN feedback is included we find a slightly
  larger fraction of stars in the BCG+ICL component.  The reason for
  this is that, although the stellar mass of the BCG+ICL components
  decreases, the total stellar mass decreases more strongly.  This
  result, in agreement with the AGN simulations by
  \citet{Puchwein2010}, is due to the combination of two different
  effects: while the AGN heating truncates the star formation at high
  redshift, mergers keep taking place and unbinding stars from
  galaxies into the diffuse component, then resulting in a net
  increase of the fraction of stars that end up in the ICL component.

As for the comparison with observational data, 
our results confirm a decreasing trend with cluster mass 
of the fraction of stars contributed by BCG and ICL, although
they predict values of this fraction that are larger than those reported by  
\citet{Gonzalez2007}, especially for massive clusters.  

\item Both of our radiative simulations show that the gas mass
  fraction within clusters increases with increasing cluster mass,
  \citep[see also ][]{Kravtsov2005, fabjan_etal10, Puchwein2010}.  Our
  results on the values of $f_{g}$ at the scale of rich and poor
  clusters, especially in the presence of AGN feedback, are in line
  with some observational data sets 
  \citep[e.g.,][]{Vikhlinin2006, Arnaud2007, Sun2009, Zhang2011}  
  suggesting that low-mass systems have proportionally less gas than
  high-mass systems.

\item The results of our analysis can be used to set priors on the
  parameters which determine the amount of gas depletion when deriving
  cosmological parameters from observations of the baryon fraction in
  massive clusters.  The baryon depletion, $Y_{\rm b}$, regardless of
  the considered radius or physics, is nearly constant with redshift,
  at least since redshift $z=1$.  However, whereas the obtained
  evolution for $Y_b$ within $R_{500}$ is virtually independent of the
  physics included, it shows some dependence on such physical processes
  when looking into inner cluster regions ($R_{2500}$).

\item Our results allow us to set a rather strong prior on the
  parameter describing the evolution of $Y_b$ (see Eq. \ref{eq:fit}),
  with $-0.02 \leq \alpha_{Y_b} \leq 0.07$, for a conservative range
  of variation holding both at $R_{500}$ and $R_{2500}$, accounting
  for the difference between different physical models and for the
  uncertainties in the estimate of the mean associated to the measured
  intrinsic scatter. As for the normalization, a conservative allowed
  range of variation can be taken to be  $Y_{0,b}=0.81\pm 0.06$ 
  at $R_{2500}$, with a slightly larger normalization and narrower uncertainty,
  $0.85\pm0.03$, at $R_{500}$.

\item We have analysed the dependences of the gas and baryonic
  depletions as a function of any combination of redshift $z$, cluster
  mass $M$ and overdensity $\Delta$, according to the functional form
  of Eq. \ref{eq:fit_extra}. In general, we find no significant
  dependences of the gas and baryonic depletions on the above
  quantities. However, we point out that caution must be taken when
  using these results for cosmological applications, given that 
  our simulated models may not span the entire range of
  models allowed by our current understanding of the intra-cluster
  medium. 
  
\end{itemize}

In general, our results show that the star formation in our radiative
runs without AGN heating, even in the presence of rather strong
galactic winds, is still too efficient, especially in small clusters
and groups.  The situation is significantly improved when AGN feedback
is included, being able to partly prevent overcooling in central
cluster regions.  However, a number of discrepancies between simulated
and observed baryonic mass fractions within clusters still exist,
especially when comparing stellar mass fractions.  Nevertheless, as
already reported by other authors \citep[e.g.,][]{fabjan_etal10}, we
can infer from our results that a feedback source associated to gas
accretion onto super-massive BHs seems to go in the right direction to
conciliate simulations with observations.

Overall, even when the real picture is far more complicated, with a
number of complex physical processes cooperating to make AGN feedback
a self-regulated process, we point out that the AGN feedback
prescription used in the present work significantly improves previous
results on the baryon census within clusters and brings closer
simulations and observations.

In general our results highlight that a robust calibration of the
baryon bias can be defined from simulations at $R_{500}$, which is
quite constant within the range of physical models for the ICM
included in our simulations. This result does not extend at the
smaller radius $R_{2500}$, which is the typical radius within which
precise measurements for the gas mass fraction have been carried out
so far for distant clusters using Chandra data \citep[][ and
references therein]{Allen2011}. While being beyond the reach of the
current generation of X--ray telescopes, tracing the gas content of
galaxy clusters out to large radii requires the next generation of
X--ray telescopes to be characterized at the same time by a large
collecting area and an excellent control of the background.

\section*{ACKNOWLEDGEMENTS}  
The authors would like to thank Volker Springel for making available
to us the non--public version of the {\small GADGET--3} code, Annalisa
Bonafede for her help with generating the initial conditions for the
simulations, Weiguang Cui for useful discussions, Gabriel Pratt for 
supplying us with observational data to compare with, and the referee
for his/her constructive comments that helped improving the
presentation of the results.  
Simulations have been carried out at
the CINECA supercomputing Centre in Bologna, with CPU time assigned
through ISCRA proposals and through an agreement with University of
Trieste.  SP acknowledges a fellowship from the European Commission's
Framework Programme 7, through the Marie Curie Initial Training
Network CosmoComp (PITN-GA-2009-238356).  DF acknowledges founding
from the Centre of Excellence for Space Sciences and Technologies
SPACE-SI, an operation partly financed by the European Union, European
Regional Development Fund and Republic of Slovenia, Ministry of Higher
Education, Science and Technology.  This work has been supported by
the PRIN-INAF09 project ``Towards an Italian Network for Computational
Cosmology'', by the PRIN-MIUR09 ``Tracing the growth of structures in
the Universe'', and by the PD51 INFN grant.  This work has been
partially supported by {\it Spanish Ministerio de Ciencia e
Innovaci\'on} (MICINN) (grants AYA2010-21322-C03-02 and
CONSOLIDER2007-00050). 

\bibliographystyle{mnbst}
\bibliography{sim_fgas}

\appendix

\begin{table*}
  \caption{Best--fit functional forms for the baryon ($f_b$), gas
    ($f_g$) and stellar ($f_{*}$) mass fractions  
    as a function of the total cluster mass, $M_{500}$, for different
    analyses of observational data.}
\label{t:fit}
\hspace*{0.3cm}{\footnotesize }
\input{tables_3.tex}
\label{tab:fittings}
\end{table*}

\begin{table*}
  \caption{Values of the gas, stellar and baryonic depletion factors 
    ($Y_{g}$, $Y_{*}$ and $Y_b$, respectively) for our set of simulated clusters,
    for the three different physical models
    ({\tt \nr}, {\tt \w}, and {\tt \agn}), computed at  
    $R_{2500}$ and $R_{500}$. In each case, values computed at redshifts
    $z=0, 0.3, 0.5, 0.8$ and 1 are reported. 
    We show within brackets the values of the intrinsic scatters computed
    within the ensamble of simulated clusters. Results are shown for
    the subset of clusters that, at each redshift, are more massive
    than $M_{500}=2\times 10^{14}\msun$. }
\input{tables_1.tex}
\label{tab:fgas}
\end{table*}

\begin{table*}
  \caption{Best--fit values of the parameters describing the evolution
    of the gas and baryonic depletions, according to Eq.~\ref{eq:fit}.
    For each simulation set ({\tt \nr}, {\tt \w}, and {\tt \agn}) and radius of interest
    ($R_{2500}$, $R_{500}$, $R_{200}$, and $R_{vir}$), we show the 
    normalization ($Y_{0,i}$) and slope ($\alpha_{Y_i}$) of the
    relation, along with their respective standard deviations within brackets, 
    as obtained from the $\chi^2$ minimization procedure.} 
  \input{tables_2.tex}
\label{tab:ajustes}
\end{table*}

\begin{table*}
  \caption{Best--fit values of the parameters describing the evolution
    of the gas and baryonic depletions according to Eq.~\ref{eq:fit_extra}.
    For each simulation set ({\tt \nr}, {\tt \w}, and {\tt \agn}) and using the data 
    within different radii of interest ($R_{2500}$, $R_{500}$ and $R_{200}$), 
    we show the best-fit values for the different parameters describing all the
    dependences of the gas and baryonic depletions
    ($Y_{0,i}$, $\alpha_{Y_i}$, $\beta_{Y_i}$, $\gamma_{Y_i}$), 
    along with their respective uncertainties within brackets.
    The quoted uncertainties on the best-fit parameters are obtained from  
    the un-weighted least-squares fit rescaled by $(\chi^2 / {\rm DOF})^{1/2}$ 
    under the assumption that the value of the true reduced $\chi^2$ is unity.
    For each fit we have also considered the case in which there is no redshift evolution, 
    i.e., $\alpha_{Y_i}$ is fixed to 0. This case is shown in the second row for each 
    simulation.
    } 
  \input{tables_extra.tex}
\label{tab:ajustes_extra}
\end{table*}

\end{document}

%% file: tables_3.tex
\begin{tabular}{ll}
  \hline 
  Sample & Best fit  \\
  \hline
  Lin et al. (2003)     & $f_{\rm b,500}=0.148^{+0.005}_{-0.004}(
{M_{500}}/[{3\times10^{14}{\rm M_{\odot}}}])^{(0.148\pm 0.040)}$ \\ 
 Giodini et al. (2009) & $f_{\rm b,500}=(0.123\pm0.003)(
{M_{500}}/[{2\times10^{14}{\rm M_{\odot}}}] )^{(0.09\pm 0.03)}$ \\ 
Lagan{\'a} et al. (2011) & $f_{\rm b,500}=0.117^{+0.060}_{-0.040}(
{M_{500}}/{10^{14}{\rm M_{\odot}}})^{(0.136\pm 0.028)}$ \\
\hline
Z11$+$S09 & $f_{\rm
g,500} =  (0.085\pm0.004)({M_{500}}/[{10^{14}{\rm M_{\odot}}}]
)^{(0.30\pm 0.07)}$    \\ 
V06$+$APP07$+$S09 & $f_{\rm g,500}=(0.093\pm0.002)(
{M_{500}}/[{2\times10^{14}{\rm M_{\odot}}}])^{(0.21\pm 0.03)}$ \\   
\hline 
 Lin et al. (2003)     & $f_{\rm *,500}=0.0164^{+0.0010}_{-0.0090}(
{M_{500}}/[{3\times10^{14}{\rm M_{\odot}}}])^{-(0.26\pm 0.09)}$ \\ 
 Gonzalez et al. (2007)  & $f_{\rm *,500}=(0.009\pm0.002){M_{500}}^{-(0.64\pm 0.13)}$  \\
 Lagan{\'a} et al. (2011) & $f_{\rm
*,500} = 0.029^{+0.008}_{-0.006} ({M_{500}}/[{10^{14.5}{\rm M_{\odot}}}]
)^{(-0.36\pm 0.17)}$ \\
 \hline
\end{tabular} 

%% file: tables_1.tex
\scalebox{0.95}{
\begin{tabular}{ 
 c@{\hspace{.6em}} c@{\hspace{.6em}} c@{\hspace{.6em}} c@{\hspace{.6em}}
 c@{\hspace{.6em}} c@{\hspace{.6em}} c@{\hspace{.6em}} c@{\hspace{.6em}}
 c@{\hspace{.6em}} c@{\hspace{.6em}} c@{\hspace{.6em}} c@{\hspace{.6em}}
 c@{\hspace{.6em}} c@{\hspace{.6em}} c@{\hspace{.6em}} }
\hline \\ 
Simulation & $z$ & & \multicolumn{3}{c}{$R_{2500}$} 
& & \multicolumn{3}{c}{$R_{500}$} \\ 
& & & $Y_{\rm g}$ & $Y_{\rm *}$ & $Y_{\rm b}$ & & 
 $Y_{\rm g}$ & $Y_{\rm *}$ & $Y_{\rm b}$ \\ %
\hline 
 \\ 
{\tt \nr} & 0.0 & &   0.80 (0.09)   &  $ -$  &    0.80 (0.09)  & &  0.84 (0.04)   &   $-$   &   0.84 (0.04) \\ 
{\tt \nr}  & 0.3 & &  0.81 (0.08)  &    $- $  &   0.81 (0.08)  & &  0.85 (0.03)  &    $- $  &   0.85 (0.03) \\ 
{\tt \nr}              & 0.5 & &   0.78 (0.09)   &   $-$   &   0.78 (0.09) &  & 0.86 (0.03)   &   $-$   &   0.86 (0.03) \\ 
 {\tt \nr}             & 0.8 & &   0.84 (0.05)   &   $- $  &   0.84 (0.05)  &  & 0.87 (0.03)  &    $- $  &   0.87 (0.03) \\ 
 {\tt \nr}             & 1.0 & &  0.84 (0.06)   &   $-$   &   0.84 (0.06) & & 0.86 (0.03)   &  $ -$   &   0.86 (0.03) \\
 \\ 
 {\tt \w} & 0.0 & & 0.49 (0.08) &   0.34 (0.07)  &  0.85 (0.06) & & 0.63 (0.04)  &  0.21 (0.04) &   0.85 (0.02)\\ 
 {\tt \w}  & 0.3 & &  0.48 (0.08)  &  0.34 (0.06)  &  0.84 (0.05) & & 0.63 (0.04)  &  0.21 (0.03)  &  0.86 (0.03) \\ 
 {\tt \w}   & 0.5 & &  0.48 (0.06)  &  0.32 (0.05)  &  0.83 (0.04)  & & 0.64 (0.03)  &  0.20 (0.02)  &  0.86 (0.02) \\ 
  {\tt \w}  & 0.8 & &  0.51 (0.06)  &  0.31 (0.05)  &  0.86 (0.05)  & & 0.64 (0.03) &   0.19 (0.02)  &  0.86 (0.02) \\ 
  {\tt \w} & 1.0 &  &  0.45 (0.06)  &  0.33 (0.06) &   0.82 (0.05) & & 0.63 (0.04) &   0.19 (0.03)  &  0.85 (0.02)   \\
 \\ 
 {\tt \agn} & 0.0 & & 0.55 (0.10)  &  0.21 (0.03) &   0.77 (0.09) & & 0.70 (0.05)  &   0.13 (0.02)  &  0.85 (0.04) \\ 
 {\tt \agn} & 0.3 & &  0.54 (0.08) &   0.21 (0.04) &   0.77 (0.07)  & & 0.70 (0.04)  &  0.13 (0.02)  &  0.85 (0.04) \\ 
 {\tt \agn} & 0.5 & &  0.55 (0.07) &   0.22 (0.04)  &  0.80 (0.08)  & & 0.70 (0.03)  &   0.13 (0.01)  &  0.86 (0.02) \\ 
 {\tt \agn} & 0.8 & &   0.55 (0.06)   &  0.21 (0.02)  &  0.79 (0.06)  & & 0.70 (0.03)  &   0.13 (0.01)  &  0.85 (0.03) \\ 
 {\tt \agn} & 1.0 & &   0.51 (0.07)  &  0.21 (0.05)  &  0.77 (0.09)  & &  0.68 (0.03)  &   0.13 (0.02)  &  0.84 (0.04) \\
\hline 
\end{tabular} }

%% file: tables_2.tex
\begin{tabular}{cccccc}
\hline Simulation & Radius & $Y_{0,g}$  & $\alpha_{Y_g}$ & $Y_{0,b}$  & $\alpha_{Y_b}$  \\   
\hline

{\tt \nr} & $R_{vir}$   &  0.87 (0.02)  &  0.00 (0.04)  &   0.87 (0.02)  &  0.00 (0.04)  \\ 
{\tt \nr} & $R_{200}$ &  0.86 (0.02)   &  0.00 (0.04)  &   0.86 (0.02)  &  0.00 (0.04) \\ 
{\tt \nr} &  $R_{500}$ &  0.85 (0.03)  &  0.02 (0.05)   &  0.85 (0.03)  &  0.02 (0.05) \\ 
{\tt \nr} & $R_{2500}$ & 0.79 (0.07)  &  0.07 (0.12)   &  0.79 (0.07)  &  0.07 (0.12) \\ 
\\
{\tt \w} & $R_{vir}$ &   0.70 (0.03) &  -0.03 (0.06)   &  0.87 (0.02) &  -0.01 (0.03) \\ 
{\tt \w} & $R_{200}$ &  0.68 (0.03)  &  0.00 (0.06)   &  0.86 (0.02)  & -0.01 (0.03) \\ 
{\tt \w} &  $R_{500}$ &   0.63 (0.03)  &  0.01 (0.08)   &  0.86 (0.02)  &  0.00 (0.03) \\ 
{\tt \w} & $R_{2500}$ &  0.49 (0.06)  &  -0.04 (0.18)  &   0.85 (0.05) &  -0.02 (0.08) \\ 
\\
{\tt \agn} & $R_{vir}$   &   0.76 (0.03) &  -0.04 (0.05)  &   0.87 (0.02) &  -0.01 (0.04) \\ 
{\tt \agn} & $R_{200}$ & 0.75 (0.03)  &  -0.03 (0.05)   &  0.87 (0.03) &  -0.01 (0.04) \\ 
{\tt \agn} &  $R_{500}$ &   0.71 (0.03) &  -0.03 (0.06)   &  0.85 (0.03)  &  0.00 (0.05) \\ 
{\tt \agn} & $R_{2500}$ &    0.55 (0.07) &  -0.04 (0.18)  &   0.78 (0.07)  &  0.01 (0.14) \\ 
\hline
\end{tabular}

%% file: tables_extra.tex
\begin{tabular}{ccccccccc}
\hline Simulation & $Y_{0,g}$  & $\alpha_{Y_g}$ & $\beta_{Y_g}$ & $\gamma_{Y_g}$ & $Y_{0,b}$  & $\alpha_{Y_b}$ & $\beta_{Y_b}$ & $\gamma_{Y_b}$  \\   
\hline

{\tt \nr} &  0.84 (0.01)  & 0.03   (0.01)  & 0.01 (0.01)  & -0.03  (0.01)  &  0.84 (0.01)  & 0.03   (0.01)  & 0.01 (0.01)  & -0.03  (0.01)  \\ 
{\tt \nr} &  0.84 (0.01) & 0.00   & 0.00 (0.01) & -0.03 (0.01)  &  0.84 (0.01) & 0.00   & 0.00 (0.01) & -0.03 (0.01)   \\ 
\\
{\tt \w} & 0.60  (0.01) &  0.06  (0.02)   & 0.07  (0.01) &  -0.13 (0.01)  & 0.85 (0.01) & 0.00 (0.01) & 0.01 (0.01) & -0.01 (0.01) \\ 
{\tt \w} & 0.61  (0.01) &  0.00   & 0.05 (0.01) & -0.13  (0.01)  & 0.85 (0.01) & 0.00  & 0.01 (0.01) & -0.01 (0.01) \\
\\
{\tt \agn} &  0.67 (0.01) &  0.02 (0.02)  &   0.06 (0.01) &  -0.12 (0.01) &  0.83 (0.01) & 0.03 (0.01)  & 0.03 (0.01) &  -0.04 (0.01) \\ 
{\tt \agn} &  0.68 (0.01) &  0.00  &  0.06  (0.01) & -0.12 (0.01) & 0.84 (0.01)  & 0.00 & 0.03  (0.01) & -0.04 (0.01) \\ 
\hline
\end{tabular}